\documentclass[journal]{IEEEtran}
\ifCLASSINFOpdf
\else
\fi
\usepackage{graphicx, color, cite, epsfig, dsfont, amssymb, amsmath, amsthm, amsfonts, amsbsy, mathrsfs, mathtools, delarray}
\usepackage[colorlinks, linkcolor =blue, citecolor=blue]{hyperref}

\newtheorem{theorem}{Theorem}
\newtheorem{definition}{Definition}

\newtheorem{algorithm}{Algorithm}
\newtheorem{remark}{Remark}

\begin{document}
%
\title{Q-learning for Optimal Control of Continuous-time Systems}
%
%
%

\author{Biao~Luo,
        Derong Liu,~\IEEEmembership{Fellow,~IEEE~}
        and~Tingwen~Huang
\thanks{Biao Luo and Derong Liu are with State Key Laboratory of Management and Control for Complex Systems, Institute of Automation, Chinese Academy of Sciences, Beijing 100190, P. R. China (E-mail: biao.luo@ia.ac.cn; derong.liu@ia.ac.cn).}
\thanks{Tingwen~Huang is with the Texas A\&M University at Qatar, PO Box 23874, Doha, Qatar (E-mail: tingwen.huang@qatar.tamu.edu).}%
}

%
%

\markboth{Manuscript Under Review}%
{Shell \MakeLowercase{\textit{et al.}}: Bare Demo of IEEEtran.cls for Journals}
%



\maketitle

\begin{abstract}
In this paper, two Q-learning (QL) methods are proposed and their convergence theories are established for addressing the model-free optimal control problem of general nonlinear continuous-time systems.
By introducing the Q-function for continuous-time systems, policy iteration based QL (PIQL) and  value iteration based QL (VIQL) algorithms are proposed for learning the optimal control policy from real system data rather than using mathematical system model.
It is proved that both PIQL and VIQL methods generate a nonincreasing Q-function sequence, which converges to the optimal Q-function.
For implementation of the QL algorithms, the method of weighted residuals is applied to derived the parameters update rule.
The developed PIQL and VIQL algorithms are essentially off-policy reinforcement learning approachs, where the system data can be collected arbitrary and thus the exploration ability is increased. With the data collected from the real system, the QL methods learn the optimal control policy offline, and then the convergent control policy will be employed to real system.
The effectiveness of the developed QL algorithms are verified through computer simulation.
\end{abstract}

\begin{IEEEkeywords}
Q-learning; model-free optimal control; off-policy reinforcement learning; the method of weighted residuals.
\end{IEEEkeywords}

%
\IEEEpeerreviewmaketitle

\section{Introduction}
%
%
%
%
\IEEEPARstart{Q}{-learning} (QL)
is a popular and powerful off-policy reinforcement learning (RL) method, which is a great breakthrough in RL researches \cite{bertsekas1996neuro,sutton1998reinforcement,kaelbling1996reinforcement,gosavi2009reinforcement}.
QL was proposed by Watkins \cite{watkins1989learning,watkins1992q} that can be used to optimally solve Markov decision processes (MDPs).
The major attractions of QL are its simplicity and that it allows using arbitrary sampling policies to generate the training data rather than using the policy to be evaluated.
Till present, Watkins' QL \cite{watkins1989learning,watkins1992q} has been extended and some meaningful results have been reported \cite{tsitsiklis1994asynchronous, watkins1992q,sutton2014new, tsitsiklis1997analysis, jaakkola1994convergence, maei2010gq, Even-Dar2003learning} in machine learning community. Such as, the theoretical analysis of QL was studied in \cite{tsitsiklis1994asynchronous, maei2010gq, jaakkola1994convergence, Even-Dar2003learning,gosavi2004reinforcement}.
An incremental multi-step QL \cite{peng1996incremental} was proposed, named Q($ \lambda $)-learning, which extends the one-step QL by combining it with TD($ \lambda $) returns for general $ \lambda $ in a natural way for delayed RL.
A GQ($ \lambda $) algorithm was introduced in \cite{maei2010gq}, which works to a general setting including eligibility traces and off-policy learning of temporally abstract predictions.
By using two-timescale stochastic approximation methodology, two QL algorithms \cite{bhatnagar2008new} were proposed.
Observe that these results about QL are mainly for MDPs, which is highly related to the optimal control problem in control community. Thus, it is possible and promising to introduce the basic QL framework for addressing the optimal control design problem.

For the optimal control problem in control community, it usually depends on the solution of the complicated Hamilton-Jacobi-Bellman equation (HJBE) \cite{hull2003optimal,bertsekas2005dynamic,lewis2013optimal}, which is extremely difficult and requires the accurate mathematical system model.
However, for many practical industrial systems, due to their large scale and complex manufacturing techniques, equipment and procedures, it is usually impossible to identify the accurate mathematical model for optimal control design, and thus the explicit expression of the HJBE is unavailable.
On the other hand, with the development and extensive applications of digital sensor technologies, and the availability of cheaper measurement and computing equipment, more and more system information could be extracted for direct control design.
In the past few years, the model-free optimal control problem has attracted researchers' extensive attention in control community \cite{lewis2013reinforcement, lewis2012reinforcement}, and has also brought new challenges to them.
Some model-free or partially model-free RL methods \cite{vrabie2009neural, zhang2011data, liu2012neural, jiang2012computational, dierks2012online, wang2012optimal, modaresadaptive2013adaptive, modares2014integral, wei2014adaptive, li2014integral, modares2014linear, yang2014reinforcement} have been developed for solving the optimal control design problem by using real system data.
For example, RL approaches were employed to solve linear quadratic regulator (LQR) problem \cite{jiang2012computational}, optimal tracking control problem \cite{modares2014linear} and zero-sum game problem \cite{li2014integral} of linear systems.
For nonlinear optimal control problem, Modares \textit{et al.} \cite{modares2014integral} developed an experience-replay based integral RL algorithm for nonlinear partially unknown constrained-input systems.
Zhang \textit{et al.} \cite{zhang2011data} presented a data-driven robust approximate optimal tracking control scheme for nonlinear systems, but it requires a prior model identification procedure and the approximate dynamic programming method is still model-based.
Globalized dual heuristic programming algorithms \cite{wang2012optimal, liu2012neural} were developed by using three neural networks (NNs) for estimating system dynamics, cost function and its derivatives, and control policy, where model NN construction error was considered.

Most recently, some QL techniques have been introduced for solving the optimal control problems \cite{al2007model,kim2010model,lee2012integral,palanisamy2014continuous,kiumarsi2014reinforcement}. Such as, for linear discrete-time systems, the optimal tracking control problem \cite{kiumarsi2014reinforcement} and $ H_\infty $ control problem \cite{al2007model,kim2010model} were studied with QL.
Online QL algorithms \cite{lee2012integral, palanisamy2014continuous}  were investigated for solving the LQR problem of linear continuous-time systems.
However, these works are just for simple linear optimal control problem \cite{lee2012integral,palanisamy2014continuous,kiumarsi2014reinforcement} or discrete-time systems \cite{al2007model,kim2010model,palanisamy2014continuous,kiumarsi2014reinforcement}. To the best of our knowledge, the QL method and its theories are still rarely studied for general nonlinear continuous-time systems.

In this paper, we consider the model-free optimal control problem of general nonlinear continuous-time systems, and two QL algorithms proposed: policy iteration based QL (PIQL) and value iteration based QL (VIQL).
The rest of is paper is arranged as follows. Section \ref{Sec_2} presents the problem description.
PIQL and VIQL algorithms are proposed in Sections \ref{Sec_3} and \ref{Sec_4}, which are then simplified for LQR problem in Section \ref{Sec_5}. Subsequently, the computer simulation results are demonstrated in Section \ref{Sec_6} and a brief conclusion is given in Section \ref{Sec_7}.

\textit{Notation}: $\mathbb{R}^n$ is the set of the $ n $-dimensional Euclidean space and  $ \Vert \cdot \Vert $ denotes it norm. The superscript $ T $ is used for the transpose and  $ I $ denotes the identify matrix of appropriate dimension. $ \nabla \triangleq \partial / \partial x $  denotes a gradient operator notation. For a symmetric matrix $ M, M>(\geq) 0$  means that it is a positive (semi-positive) definite matrix.  $ \Vert v \Vert ^2_M \triangleq v^T M v $  for some real vector $ v $ and symmetric matrix $ M>(\geq) 0$  with appropriate dimensions. $ C^1(\mathcal {X}) $  is a function space on $ \mathcal {X} $  with first derivatives are continuous. Let $ \mathcal{X} $ and $\mathcal{U} $  be compact sets, denote $ \mathcal {D}  \triangleq \lbrace (x,u,x') | x,x' \in \mathcal{X}, u \in \mathcal{U} \rbrace $. For column vector functions $ s_1(x,u,x') $ and $ s_2(x,u,x') $, where $ (x,u,x') \in \mathcal {D} $, define the inner product $ \langle s_1(x,u,x'), s_2(x,u,x')\rangle_{\mathcal {D} } \triangleq \int_{\mathcal {D} } s_1^T(x,u,x') s_2(x,u,x') d(x,u,x') $  and the norm  $ \Vert s_1(x,u,x') \Vert _{\mathcal {D} } \triangleq  \langle s_1(x,u,x'), s_1(x,u,x')\rangle_{\mathcal {D} }^{1/2}$.

\section{Problem Description} \label{Sec_2}
Let us consider the following general nonlinear continuous-time  system:
\begin{equation}\label{eq_2.1}
\dot{x}(t) = f(x(t), u(t)), x(0)=x_0
\end{equation}
where $x = [x_1~ ... ~x_n]^T \in \mathcal {X} \subset \mathbb{R}^n$ is the state, $x_0$ is the initial state and $u = [u_{1}~ ... ~u_{m}]^T \in \mathcal{U} \subset \mathbb{R}^m$ is the control input. Assume that $ f(x,u) $ is Lipschitz continuous on the set $ \mathcal{X} \times \mathcal{U} $ that contains the origin, i.e., $ f(0,0)=0 $. The system is stabilizable on  $\mathcal {X}$, i.e., there exists a continuous control function $ u(x) $  such that the system is asymptotically stable on  $\mathcal {X}$.

For the model-free optimal control problem considered in this paper, the model $ f(x,u) $ of system \eqref{eq_2.1} is completely \textit{unknown}. The objective of optimal control design is to find a state feedback control law $ u(t) = u(x(t)) $, such that the system \eqref{eq_2.1} is closed-loop asymptotically stable, and minimize the following generalized infinite horizon cost functional:
\begin{equation}\label{eq_2.2}
V_u(x_0) \triangleq \int_{0}^{+\infty}(S(x(t))+W(u(t)))dt
\end{equation}
where $ S(x) $ and $ W(u) $ are positive definite functions, i.e., for $ \forall x \neq 0, u \neq 0, S(x)>0, W(u)>0 $, and $ S(x)=0, W(u)=0  $ only when $ x = 0, u = 0 $. Then, the optimal control problem is briefly presented as
\begin{equation} \label{eq_2.3}
u(t) = u^*(x) \triangleq \arg\min_u V_u(x_0).
\end{equation}

\section{Policy Iteration Based Q-learning} \label{Sec_3}
In this section, a PIQL algorithm and its convergence are established for solving the model-free optimal control problem of the system \eqref{eq_2.1}. Before starting, the definition of admissible control is necessary.
\begin{definition} \label{def_3.1}
(Admissible control) For the given system \eqref{eq_2.1}, $ x \in \mathcal {X} $, a control policy $ u(x) $ is defined to be admissible with respect to cost function \eqref{eq_2.2} on $ \mathcal {X} $, denoted by  $ u(x) \in \mathfrak{U}(\mathcal {X}) $, if, 1)  $ u $ is continuous on $ \mathcal {X} $, 2)  $ u(0)=0 $, 3)  $ u(x) $ stabilizes the system, and 4)  $ V_u(x)<\infty, \forall x \in \mathcal {X} $. $ \square $
\end{definition}

\subsection{Policy Iteration Based Q-learning} \label{Sec_3.1}
Noting that the mathematical system model $ f(x,u) $ is unknown, a PIQL algorithm is proposed to learn the optimal control policy from real system data directly.
For notations simplicity, denote $ t' \triangleq t + \Delta t $ for $ \forall t ,\Delta t >0$,  $ x_t \triangleq x(t) $ and $ x'_t \triangleq x(t') $.
For an admissible control policy $ u(x) \in \mathfrak{U}(\mathcal {X}) $, define its cost function
\begin{equation} \label{eq_3.1}
V_u(x_t) \triangleq \int_{t}^{+\infty}\mathcal {R} (x(\tau),u(\tau)) d\tau
\end{equation}
where $ \mathcal {R} (x,u) \triangleq S(x)+W(u) $, and $ V_u(0) = 0 $.
Define the Hamilton function
\begin{equation} \label{eq_3.2}
H(x,u,\nabla V) \triangleq [\nabla V(x)]^T f(x,u) + \mathcal {R} (x,u)
\end{equation}
for some cost function $ V(x) \in C^1(\mathcal {X}) $.
Taking derivative on both sides of expression \eqref{eq_3.1} along with the system \eqref{eq_2.1} yields
\begin{flalign}
[\nabla V_u(x)]^T f(x,u) = -\mathcal {R} (x,u) \nonumber
\end{flalign}
i.e.,
\begin{flalign} \label{eq_3.3}
H(x,u,\nabla V_u) = 0
\end{flalign}
which is linear partial differential equation.

Let $ V^*(x) \triangleq V_{u^*}(x) $ be the optimal cost function, then the optimal control law \eqref{eq_2.3} is given by
\begin{equation} \label{eq_3.4}
u^*(x) \triangleq \arg\min_u H(x,u,\nabla V^*).
\end{equation}
Substituting \eqref{eq_3.4} into \eqref{eq_3.3} yields the following HJBE
\begin{equation} \label{eq_3.5}
H(x,u^*,\nabla V^*) = 0
\end{equation}
which is nonlinear partial differential equation. It is observed that the optimal control policy $ u^* $ relies on the solution $ V^* $ of the HJBE \eqref{eq_3.5}, while the unavailability of the system model $ f(x,u) $ prevents using model-based approaches for control design.

To derive the PIQL algorithm, it is necessary to introduce an action-state value function, named Q-function. For $ \forall u(x) \in \mathfrak{U}(\mathcal {X}) $, define its Q-function as
\begin{equation} \label{eq_3.6}
Q_u(x_t, \mu) \triangleq \int_{t}^{t'}\mathcal {R} (x(\tau),\mu(\tau)) d \tau + \int_{t'}^{+\infty} \mathcal {R} (x(\tau),u(\tau)) d\tau.
\end{equation}
where $ (x_t, \mu) \in \mathcal{X} \times \mathcal{U} $ and $ Q_u(0, 0) = 0 $. It is found that $ Q_u(x, u) = V_u(x) $, then the Q-function \eqref{eq_3.6} is rewritten as
\begin{flalign} \label{eq_3.7a}
Q_u(x_t, \mu) & = \int_{t}^{t'}\mathcal {R} (x(\tau),\mu(\tau)) d \tau + Q_u(x'_t, u) \nonumber \\
& = \int_{t}^{t'}\mathcal {R} (x(\tau),\mu(\tau)) d \tau + V_u(x'_t).
\end{flalign}
Note that the Q-function $ Q_u(x, \mu) $ for a state $ x $ and control action $ \mu $ represents the value of the performance metric obtained when action $ \mu $ is used in state $ x $ and the control policy $ u $ is pursued thereafter.
For the optimal control policy $ u^*(x) $, the associate optimal Q-function $ Q^*(x,\mu) \triangleq Q_{u^*}(x,\mu) $ is given by
\begin{flalign} \label{eq_3.7}
Q^*(x_t, \mu) & = \int_{t}^{t'}\mathcal {R} (x(\tau),\mu(\tau)) d \tau + Q_{u^*}(x'_t, u^*) \nonumber \\
& = \int_{t}^{t'}\mathcal {R} (x(\tau),\mu(\tau)) d \tau + V^*(x'_t).
\end{flalign}
Then,
\begin{flalign} \label{eq_3.8}
Q^*(x_t, u^*) & = \min_\mu Q^*(x_t, \mu) \nonumber \\
&= \min_\mu \int_{t}^{t'}\mathcal {R} (x(\tau),\mu(\tau)) d \tau + V^*(x'_t) \nonumber \\
&= \int_{t}^{t'}\mathcal {R} (x(\tau),u^*(\tau)) d \tau + V^*(x'_t) \nonumber \\
&= V^*(x_t).
\end{flalign}
According to the expressions \eqref{eq_2.3} and \eqref{eq_3.8}, the optimal control policy $ u^*(x) $ can also be presented as
\begin{equation} \label{eq_3.9}
u^*(x) = \arg\min_\mu V_u(x) =\arg\min_\mu Q^*(x, \mu).
\end{equation}

Now, we give the PIQL method as follows:

\noindent \rule{0.49\textwidth}{2pt}
\begin{algorithm} \label{algorithm_3.1}
\emph{
Policy iteration based Q-learning\\
\noindent \rule{0.49\textwidth}{1pt}
\begin{itemize}
\item [$\blacktriangleright$] \emph{Step 1:}
Let $ u^{(0)}(x) \in \mathfrak{U}(\mathcal {X}) $ be an initial control policy, and $ i = 0 $;
\item [$\blacktriangleright$] \emph{Step 2:}
(\textbf{Policy evaluation}) Solve the equation
\begin{equation} \label{eq_3.10}
Q^{(i)}(x_t, \mu) = \int_{t}^{t'}\mathcal {R} (x(\tau),\mu(\tau)) d \tau + Q^{(i)}(x'_t, u^{(i)})
\end{equation}
for unknown Q-function $ Q^{(i)} \triangleq Q_{u^{(i)}} $;
\item [$\blacktriangleright$] \emph{Step 3:}
(\textbf{Policy improvement}) Update control policy with
\begin{equation} \label{eq_3.11}
u^{(i+1)}(x) = \arg\min_\mu Q^{(i)}(x, \mu);
\end{equation}
\item [$\blacktriangleright$] \emph{Step 4:}
Let $ i = i+1 $, go back to Step 2 and continue.
$\square $
\end{itemize}
}
\end{algorithm}
\noindent \rule{0.49\textwidth}{2pt}
\begin{remark} \label{remark1}
\emph{
The PIQL Algorithm \ref{algorithm_3.1} involves two basic operators: policy evaluation and policy improvement. The policy evaluation is to evaluate the current control policy $ u^{(i)} $ for its Q-function $ Q^{(i)} $, and the policy improvement is to get a better control policy $  u^{(i+1)} $ based on the Q-function $ Q^{(i)} $.
By giving an initial admissible control policy, PIQL algorithm implement policy evaluation and policy improvement alternatively to learn the optimal Q-function $ Q^* $ and the optimal control policy $ u^* $.
Note that the proposed PIQL algorithm has two main features. First, it is a data-based control design approach, where the system model $ f(x,u) $ is not required.
Second, it is an off-policy learning approach  \cite{sutton1998reinforcement,precup2001off,luo2014off}, which refers to evaluate a target policy $ u^{(i)} $ for its Q-function $ Q^{(i)} $  while following another policy $ \mu $, known as behavior policy that can be exploratory. Thus, the proposed PIQL algorithm is exploration insensitive, which means that it is independent of how the behaves while the data is being collected.
$ \square $ }
\end{remark}

\subsection{Theoretical Analysis} \label{Sec_3.2}
Some theoretical aspects of the PIQL method (i.e., Algorithm \ref{algorithm_3.1}) are analyzed in this subsection. Its convergence is proved by demonstrating that the sequences $ \{ Q^{(i)} \} $ and $ \{ u^{(i)} \} $ generated by Algorithm \ref{algorithm_3.1} will converge to the optimal Q-function $ Q^* $ and the optimal control policy $ u^* $, respectively.
\begin{theorem}\label{theorem_3.1}
Let $ u^{(i+1)}(x) $ be given by \eqref{eq_3.11},  then
\begin{equation} \label{eq_3.12}
u^{(i+1)}(x) = \arg\min_\mu H(x,\mu,\nabla V^{(i)}).
\end{equation}
\end{theorem}
\noindent \textbf{Proof.} According to \eqref{eq_3.7a}, equation \eqref{eq_3.10} can be rewritten as
\begin{equation} \label{eq_3.13}
Q^{(i)}(x_t, \mu) = \int_{t}^{t'}\mathcal {R} (x(\tau),\mu(\tau)) d \tau + V^{(i)}(x'_t)
\end{equation}
where $ V^{(i)}(x'_t) \triangleq V_{u^{(i)}}(x'_t) = Q^{(i)}(x'_t, u^{(i)}) $.
Subtracting $ V^{(i)}(x_t) $ on both sides of equation \eqref{eq_3.13} yields
\begin{flalign}
& Q^{(i)}(x_t, \mu) - V^{(i)}(x_t)  \nonumber \\
& = \int_{t}^{t'}\mathcal {R} (x(\tau),\mu(\tau)) d \tau + V^{(i)}(x'_t) - V^{(i)}(x_t) \nonumber \\
& = \int_{t}^{t'}\mathcal {R} (x(\tau),\mu(\tau)) d \tau
+ \int_{t}^{t'} \frac{dV^{(i)}(x)}{d \tau} d \tau \nonumber \\
& = \int_{t}^{t'} \left( \mathcal {R} (x(\tau),\mu(\tau)) +  [\nabla V^{(i)} (x(\tau))]^T f(x(\tau),\mu(\tau)) \right) d \tau \nonumber \\
& = \int_{t}^{t'}  H(x(\tau),\mu(\tau),\nabla V^{(i)} (x(\tau))) d \tau \nonumber
\end{flalign}
i.e.,
\begin{equation}
Q^{(i)}(x_t, \mu) = V^{(i)}(x_t) + \int_{t}^{t'}  H(x(\tau),\mu(\tau),\nabla V^{(i)} (x(\tau))) d \tau. \nonumber
\end{equation}
Then, 
\begin{flalign} \label{eq_3.14}
& \min_\mu Q^{(i)}(x_t, \mu) = \min_\mu V^{(i)}(x_t) \nonumber \\
& \quad + \min_\mu \int_{t}^{t'}  H(x(\tau),\mu(\tau),\nabla V^{(i)} (x(\tau))) d \tau  \nonumber \\
& = V^{(i)}(x_t) + \min_\mu \int_{t}^{t'}  H(x(\tau),\mu(\tau),\nabla V^{(i)} (x(\tau))) d \tau.
\end{flalign}

Let
\begin{equation} \label{eq_3.15}
\mu_1(\tau) = \arg\min_\mu \int_{t}^{t'}  H(x(\tau),\mu(\tau),\nabla V^{(i)} (x(\tau))) d \tau
\end{equation}
and
\begin{equation} \label{eq_3.16}
\mu_2(\tau) = \arg\min_\mu H(x(\tau),\mu(\tau),\nabla V^{(i)} (x(\tau))).
\end{equation}
Thus, it follows from \eqref{eq_3.16} that
\begin{flalign}
H(x,\mu_2(\tau),\nabla V^{(i)} (x))
&= \min_\mu H(x,\mu,\nabla V^{(i)} (x)) \nonumber \\
&\leqslant  H(x,\mu_1(\tau),\nabla V^{(i)} (x))  \nonumber
\end{flalign}
for $ \forall \tau \in [t, t'] $,
then integrating on interval $ [t,t'] $ yields
\begin{flalign} \label{eq_3.17}
& \int_{t}^{t'}  H(x(\tau),\mu_2(\tau),\nabla V^{(i)} (x(\tau))) d \tau \nonumber \\
& \quad \leqslant \int_{t}^{t'}  H(x(\tau),\mu_1(\tau),\nabla V^{(i)} (x(\tau))) d \tau.
\end{flalign}
On the other hand, based on \eqref{eq_3.15}, 
\begin{flalign} \label{eq_3.18}
& \int_{t}^{t'}  H(x(\tau),\mu_1(\tau),\nabla V^{(i)} (x(\tau))) d \tau \nonumber \\
& \quad =  \min_\mu \int_{t}^{t'}  H(x(\tau),\mu(\tau),\nabla V^{(i)} (x(\tau))) d \tau \nonumber \\
& \quad \leqslant \int_{t}^{t'}  H(x(\tau),\mu_2(\tau),\nabla V^{(i)} (x(\tau))) d \tau.
\end{flalign}

According to \eqref{eq_3.17} and \eqref{eq_3.18}, 
\begin{flalign}
&\int_{t}^{t'}  H(x(\tau),\mu_1(\tau),\nabla V^{(i)} (x(\tau))) d \tau \nonumber \\
&\quad = \int_{t}^{t'}  H(x(\tau),\mu_2(\tau),\nabla V^{(i)} (x(\tau))) d \tau \nonumber
\end{flalign}
i.e.,
\begin{flalign} \label{eq_3.19}
&\min_\mu \int_{t}^{t'}  H(x(\tau),\mu(\tau),\nabla V^{(i)} (x(\tau))) d \tau \nonumber \\
&\quad = \int_{t}^{t'} \min_\mu  H(x(\tau),\mu(\tau),\nabla V^{(i)} (x(\tau))) d \tau.
\end{flalign}
Then, it follows from equations \eqref{eq_3.14} and \eqref{eq_3.19} that
\begin{flalign} \label{eq_3.20}
& \min_\mu Q^{(i)}(x_t, \mu) = V^{(i)}(x_t) \nonumber \\
& \quad + \int_{t}^{t'}  \min_\mu  H(x(\tau),\mu(\tau),\nabla V^{(i)} (x(\tau))) d \tau.
\end{flalign}
which means that
\begin{equation} \label{eq_3.21}
\arg\min_\mu Q^{(i)}(x, \mu) = \arg\min_\mu H(x,\mu,\nabla V^{(i)}) = u^{(i+1)}(x) .
\end{equation}
The proof is completed. $ \square $

From Theorem \ref{theorem_3.1}, it is indicated that the policy improvement with \eqref{eq_3.11} for learning a control policy $ u^{(i+1)} $ that minimizes the Q-function $ Q^{(i)} $, is theoretically equivalent to obtain a control policy that minimizes the associated Hamilton function $ H(x,\mu,\nabla V^{(i)}) $.

\begin{theorem}\label{theorem_3.2}
Let $ u^{(0)}(x) \in \mathfrak{U}(\mathcal {X}) $, the sequence $ \{ u^{(i)}(x) \} $ be generated by Algorithm \ref{algorithm_3.1}. Then, $ u^{(i)}(x) \in \mathfrak{U}(\mathcal {X}) $ for $ \forall i = 0,1,2,... $.
\end{theorem}
\noindent \textbf{Proof.} The proof is by mathematical induction. First, $ u^{(0)}(x) \in \mathfrak{U}(\mathcal {X}) $, i.e., Theorem \ref{theorem_3.2} holds for $ i=0 $.

Assume that Theorem \ref{theorem_3.2} holds for index $ i = l$, that is, $ u^{(l)}(x) \in \mathfrak{U}(\mathcal {X}) $. Then, according to \eqref{eq_3.3}, the cost function $ V^{(l)}(x) $ associated with $ u^{(l)}(x) $ satisfies the following Hamilton function equation
\begin{equation}\label{eq_3.22}
H(x,u^{(l)},\nabla V^{(l)}) = 0.
\end{equation}
Next, we should prove that Theorem \ref{theorem_3.2} still holds for index $ i = l+ 1 $. Under the control policy $ u^{(l+1)}(x) $, the closed-loop system is
\begin{equation}\label{eq_3.23}
\dot{x} = f(x, u^{(l+1)}).
\end{equation}
Selecting $ V^{(l)}(x) $ be the Lyapunov function, and taking derivative of $ V^{(l)}(x) $ along the state of the closed-loop system \eqref{eq_3.23} yields
\begin{flalign} \label{eq_3.24}
\dot{V}^{(l)} =&  \nabla V^{(l)} f(x, u^{(l+1)}) \nonumber \\
=&  \nabla V^{(l)} f(x, u^{(l+1)}) + S(x) + W(u^{(l+1)})  \nonumber \\
&-  S(x) - W(u^{(l+1)}) \nonumber \\
=&  H(x, u^{(l+1)}, \nabla V^{(l)}) -  S(x) - W(u^{(l+1)}).
\end{flalign}
Based on the expressions \eqref{eq_3.12} and \eqref{eq_3.22}, 
\begin{flalign} \label{eq_3.25}
H(x, u^{(l+1)}, \nabla V^{(l)}) & = \min_\mu H(x,\mu,\nabla V^{(l)}) \nonumber \\
& \leqslant H(x, u^{(l)}, \nabla V^{(l)}) \nonumber \\
& = 0.
\end{flalign}
It follows from \eqref{eq_3.24} and \eqref{eq_3.25} that
\begin{equation}
\dot{V}^{(l)} \leqslant -  S(x) - W(u^{(l+1)}) \leqslant 0. \nonumber
\end{equation}
which means that the closed-loop system \eqref{eq_3.23} is asymptotically stable. Thus, $ u^{(l+1)}(x) \in \mathfrak{U}(\mathcal {X}) $, i.e., Theorem \ref{theorem_3.2} holds for index $ i = l+1 $.
$ \square $.

Theorem \ref{theorem_3.2} shows that giving an initial admissible control policy, all policies generated by the PIQL Algorithm \ref{algorithm_3.1} are admissible.

\begin{theorem}\label{theorem_3.3}
For $ \forall (x, \mu) \in \mathcal{X} \times \mathcal{U} $, the sequences $ \{ Q^{(i)}(x,\mu) \} $ and $ \{ u^{(i)}(x) \} $ are generated by Algorithm \ref{algorithm_3.1}. Then, \\
1) $ Q^{(i)}(x,\mu) \geqslant Q^{(i+1)}(x,\mu) \geqslant Q^*(x,\mu) $ \\
2) $ Q^{(i)}(x,\mu) \rightarrow Q^*(x,\mu) $ and $ u^{(i)}(x) \rightarrow u^*(x) $ as $ i \rightarrow \infty $.
\end{theorem}
\noindent \textbf{Proof.} 1) For $ \forall i = 0,1,2,... $, define a new iterative function sequence
\begin{equation}\label{eq_3.26}
\overline{V}^{(i)}(x_t) = \min_\mu Q^{(i)}(x_t,\mu).
\end{equation}
Then, it follows from the expressions \eqref{eq_3.10} and \eqref{eq_3.26}
 that
\begin{flalign} \label{eq_3.27}
\overline{V}^{(i)}(x_t) & \leqslant Q^{(i)}(x_t,u^{(i)}) \nonumber \\
& = \int_{t}^{t'}\mathcal {R} (x(\tau),u^{(i)}(\tau)) d \tau + V^{(i)}(x'_t) \nonumber \\
& = V^{(i)}(x_t).
\end{flalign}

For $ \forall x_t \in \mathcal {X} $, according to \eqref{eq_3.1}, 
\begin{flalign} \label{eq_3.28}
V^{(i+1)}(x_t)
= & \int_{t}^{t + \Delta t}\mathcal {R} (x(\tau),u^{(i+1)}(\tau)) d \tau + V^{(i+1)}(x'_t) \nonumber \\
= & \int_{t}^{t + \Delta t}\mathcal {R} (x(\tau),u^{(i+1)}(\tau)) d \tau + V^{(i)}(x'_t) \nonumber \\
& - V^{(i)}(x'_t) + V^{(i+1)}(x'_t)  \nonumber \\
=&  \overline{V}^{(i)}(x_t) - V^{(i)}(x'_t) + V^{(i+1)}(x'_t) \nonumber \\
\leqslant & V^{(i)}(x_t) - V^{(i)}(x'_t) + V^{(i+1)}(x'_t) \nonumber \\
=  & \int_{t}^{t + \Delta t}\mathcal {R} (x(\tau),u^{(i)}(\tau)) d \tau + V^{(i+1)}(x'_t).
\end{flalign}
Similar with \eqref{eq_3.28}, there is
\begin{flalign}\label{eq_3.29}
V^{(i+1)}(x_{t + k\Delta t})  \leqslant & \int_{t + k\Delta t}^{t + (k+1)\Delta t}\mathcal {R} (x(\tau),u^{(i)}(\tau)) d \tau \nonumber \\
& + V^{(i+1)}(t + x_{(k+1)\Delta t})
\end{flalign}
for $ \forall k $. Note from Theorem \ref{theorem_3.2} that $ u^{(i+1)}(x) \in \mathfrak{U}(\mathcal {X}) $, then $ x_t = 0 $ as $ t \rightarrow \infty $, that is $ V^{(i+1)}(x_{t \rightarrow \infty}) = 0 $. Thus, it follows from \eqref{eq_3.28} and \eqref{eq_3.29} that
\begin{flalign} \label{eq_3.30}
V^{(i+1)}(x_t)
 \leqslant &  \int_{t}^{t + \Delta t}\mathcal {R} (x(\tau),u^{(i)}(\tau)) d \tau \nonumber \\
& +  \int_{t + \Delta t}^{t + 2\Delta t}\mathcal {R} (x(\tau),u^{(i)}(\tau)) d \tau \nonumber \\
& + V^{(i+1)}(x_{t + 2\Delta t}) \nonumber \\
\leqslant & \int_{t}^{t + \Delta t}\mathcal {R} (x(\tau),u^{(i)}(\tau)) d \tau \nonumber \\
& +  \int_{t + \Delta t}^{t + 2\Delta t}\mathcal {R} (x(\tau),u^{(i)}(\tau)) d \tau \nonumber \\
& + \cdots + \int_{t + k\Delta t}^{t + (k+1)\Delta t}\mathcal {R} (x(\tau),u^{(i)}(\tau)) d \tau  \nonumber \\
& + \cdots +  V^{(i+1)}(x_{t \rightarrow \infty}) \nonumber \\
\leqslant & \int_{t}^{\infty}\mathcal {R} (x(\tau),u^{(i)}(\tau)) d\tau \nonumber \\
 = & V^{(i)}(x_t).
\end{flalign}

For $ \forall (x_t, \mu) \in \mathcal{X} \times \mathcal{U} $, by using the expressions \eqref{eq_3.10} and \eqref{eq_3.30}, there is
\begin{flalign}
Q^{(i+1)}(x_t, \mu) = & \int_{t}^{t'}\mathcal {R} (x(\tau),\mu(\tau)) d \tau + V^{(i+1)}(x'_t)\nonumber \\
\leqslant & \int_{t}^{t'}\mathcal {R} (x(\tau),\mu(\tau)) d \tau + V^{(i)}(x'_t)\nonumber \\
= & Q^{(i)}(x_t, \mu). \nonumber
\end{flalign}

According to \eqref{eq_3.7a} and \eqref{eq_3.7}, 
\begin{flalign}
Q^{(i)}(x_t, \mu)
= & \int_{t}^{t'}\mathcal {R} (x(\tau),\mu(\tau)) d \tau + V^{(i)}(x'_t) \nonumber \\
\geqslant & \int_{t}^{t'}\mathcal {R} (x(\tau),\mu(\tau)) d \tau + V^*(x'_t) \nonumber \\
= & Q^*(x_t, \mu). \nonumber
\end{flalign}
The part 1) of Theorem \ref{theorem_3.3} is proved.

2) From the part 1) of Theorem \ref{theorem_3.3}, $ \{ Q^{(i)}(x,\mu) \} $ is a nonincreasing sequence and bounded below by $ Q^*(x,\mu)  $. Considering that a bounded monotone sequence always has a limit, denote $ Q^{(\infty)}(x,\mu) \triangleq  \lim_{i \rightarrow \infty} Q^{(i)}(x,\mu) $ and then $ u^{(\infty)}(x) \triangleq \arg\min_\mu Q^{(\infty)}(x, \mu) $.

Based on the proof of the part 1) in Theorem \ref{theorem_3.3}, $ \{ V^{(i)}(x) \} $ is also a nonincreasing sequence and bounded below by $ V^*(x) $.
Denote $ V^{(\infty)}(x) \triangleq  \lim_{i \rightarrow \infty} V^{(i)}(x) $.
It follows from Theorem \ref{theorem_3.1} (by simplify replacing $ V^{(i)} $ with $ V^{(\infty)} $) that
\begin{equation} \label{eq_3.31}
u^{(\infty)}(x) = \arg\min_\mu Q^{(\infty)}(x, \mu)
= \arg\min_\mu H(x,\mu,\nabla V^{(\infty)}).
\end{equation}
Since $ u^{(\infty)}(x) \in \mathfrak{U}(\mathcal {X}) $, it is based on \eqref{eq_3.3} that
$
H(x,u^{(\infty)},\nabla V^{(\infty)}) = 0
$,
which means that $ V^{(\infty)}(x) $ satisfies the HJBE \eqref{eq_3.5}. According to the uniqueness of the HJBE's solution, $ V^{(\infty)}(x) =  V^*(x) $.

Then, from \eqref{eq_3.10}, 
\begin{flalign} \label{eq_3.32}
Q^{(\infty)}(x_t,\mu) = &  \lim_{i \rightarrow \infty} Q^{(i)}(x_t,\mu)  \nonumber \\
=  & \lim_{i \rightarrow \infty} \int_{t}^{t'}\mathcal {R} (x(\tau),\mu(\tau)) d \tau +  \lim_{i \rightarrow \infty} Q^{(i)}(x'_t, u^{(i)}) \nonumber \\
= & \int_{t}^{t'}\mathcal {R} (x(\tau),\mu(\tau)) d \tau + Q^{(\infty)}(x'_t,u^{(\infty)}) \nonumber \\
= & \int_{t}^{t'}\mathcal {R} (x(\tau),\mu(\tau)) d \tau + V^{(\infty)}(x'_t) \nonumber \\
= & \int_{t}^{t'}\mathcal {R} (x(\tau),\mu(\tau)) d \tau + V^*(x'_t) \nonumber \\
= & Q^*(x_t,\mu).
\end{flalign}
The substitution of \eqref{eq_3.32} into \eqref{eq_3.31} yields
$ u^{(\infty)}(x) = \arg\min_\mu Q^*(x, \mu) = u^*(x) $. The proof is completed.
$ \square $

In Theorem \ref{theorem_3.3}, the convergence of the proposed PIQL algorithm is proved. It is demonstrated that $ \{ Q^{(i)}(x,\mu) \} $ is a bounded nonincreasing sequence that converges to the optimal Q-function $ Q^*(x,\mu) $, and then the control sequence $ \{ u^{(i)}(x) \} $ converges to the optimal control policy $ u^*(x) $.

\subsection{The Method of Weighted Residuals} \label{Sec_3.3}
In the PIQL algorithm (i.e., Algorithm \ref{algorithm_3.1}), the policy evaluation requires the solution of the equation \eqref{eq_3.10} for unknown Q-function $ Q^{(i)}(x,\mu) $. In this subsection, the method of weighted residuals (MWR) is developed. Let $ \Psi (x,\mu) \triangleq \{ \psi_j (x,\mu) \}_{j=1}^\infty $ be complete set of linearly independent basis functions, such that $ \psi_j (0,0) = 0 $ for $ \forall j$.
Then, the solution $ Q^{(i)}(x,\mu) $ of the iterative equation \eqref{eq_3.10} can be expressed as linear combination of basis function set $ \Psi (x,\mu) $, i.e., $ Q^{(i)}(x,\mu) = \sum_{j=1}^{\infty} \theta_{j}^{(i)} \psi_j(x,\mu) $ which are assumed to converge pointwise in $ \mathcal{X} \times \mathcal{U} $. The trial solution for $ Q^{(i)}(x,\mu) $ can be respectively taken by truncating the series to
\begin{equation}\label{eq_3.33}
\hat{Q}^{(i)} (x,\mu) = \sum_{j=1}^{L} \theta_{j}^{(i)} \psi_j(x,\mu) = \Psi_L^T(x,\mu) \theta^{(i)}
\end{equation}
where  $\theta^{(i)} \triangleq [\theta_{1}^{(i)}~...~\theta_{L}^{(i)}]^T$ is an unknown weight vector, $\Psi_L(x, \mu) \triangleq [\psi_1(x,\mu)~...~\psi_{L}(x,\mu)]^T$.
By using \eqref{eq_3.11} and \eqref{eq_3.33}, the estimated solution for $ Q^{(i)}(x,\mu) $ is given by
\begin{equation} \label{eq_3.34}
\hat{u}^{(i)}(x) = \arg\min_\mu \hat{Q}^{(i-1)}(x, \mu).
\end{equation}

Due to the truncation error of the trail solution \eqref{eq_3.33}, the replacement of $ Q^{(i)} (x,\mu) $ and  $ u^{(i)}(x) $ in the iterative equation \eqref{eq_3.10} with $\hat{Q}^{(i)} (x,\mu)$ and  $ \hat{u}^{(i)}(x) $ respectively, yields the following residual error:
\begin{flalign}\label{eq_3.35}
\sigma ^{(i)} (x_t, \mu, x'_t) \triangleq &
\hat{Q}^{(i)}(x_t, \mu) - \hat{Q}^{(i)}(x'_t, \hat{u}^{(i)}) \nonumber\\
& - \int_{t}^{t'}\mathcal {R} (x(\tau),\mu(\tau)) d \tau \nonumber\\
= & [\Psi_L(x_t,\mu) - \Psi_L(x'_t,\hat{u}^{(i)}) ]^T \theta^{(i)} \nonumber\\
& - \int_{t}^{t'}\mathcal {R} (x(\tau),\mu(\tau)) d \tau \nonumber\\
= & \overline{\rho}^{(i)}(x_t, \mu, x'_t) \theta^{(i)} - \pi(x_t, \mu)
\end{flalign}
where  $ \overline{\rho}^{(i)}(x_t, \mu, x'_t) = [\overline{\rho}_1^{(i)}(x_t, \mu, x'_t)~\cdots~\overline{\rho}_L^{(i)}(x_t, \mu, x'_t)] $ and $ \pi(x_t, \mu) \triangleq \int_{t}^{t'}\mathcal {R} (x(\tau),\mu(\tau)) d \tau $, with $ \overline{\rho}_j^{(i)}(x_t, \mu, x'_t) =  \psi_j(x_t,\mu) - \psi_j(x'_t,\hat{u}^{(i)}), j = 1,...,L $.
In the MWR, the unknown constant vector $ \theta^{(i)}  $ can be solved in such a way that the residual error $ \sigma ^{(i)} (x_t, \mu, x'_t) $  (for $ \forall t, t' \geqslant 0 $) of \eqref{eq_3.35} is forced to be zero in some average sense. The weighted integrals of the residual are set to zero:
\begin{equation} \label{eq_3.36}
\left<\mathcal{W}^{(i)}_L(x, \mu, x'), \sigma ^{(i)} (x, \mu, x') \right> _{\mathcal {D} } = 0.
\end{equation}
where $ \mathcal{W}^{(i)}_L(x, \mu, x') \triangleq  [\omega^{(i)}_1(x, \mu, x')~~\cdots~~\omega^{(i)}_L(x, \mu, x')]^T$ is named the weighted function vector. Then, the substitution of \eqref{eq_3.35} into \eqref{eq_3.36} yields,
\begin{flalign}
& \left< \mathcal{W}^{(i)}_L(x, \mu, x'), \overline{\rho}^{(i)}(x, \mu, x') \right> _{\mathcal {D} } \theta^{(i)} \nonumber \\
& \qquad -  \left<\mathcal{W}^{(i)}_L(x, \mu, x'),  \pi(x,\mu) \right>_{\mathcal {D} } = 0 \nonumber
\end{flalign}
where the notations $ \left< \mathcal{W}^{(i)}_L, \overline{\rho}^{(i)} \right> _{\mathcal {D} } $ and $ \left<\mathcal{W}^{(i)}_L,  \pi \right> _{\mathcal {D} } $ are given by
\begin{equation}
\left< \mathcal{W}^{(i)}_L, \overline{\rho}^{(i)} \right> _{\mathcal {D} } \triangleq
\left[ \begin{array}{ccc}
 		\left< \omega^{(i)}_1, \overline{\rho}_1^{(i)} \right> _{\mathcal {D} } & \cdots & \left< \omega^{(i)}_1, \overline{\rho}_L^{(i)} \right> _{\mathcal {D} } \\
 		\vdots & \cdots & \vdots \\
 		\left< \omega^{(i)}_L, \overline{\rho}_1^{(i)} \right> _{\mathcal {D} } & \cdots & \left< \omega^{(i)}_L, \overline{\rho}_L^{(i)} \right> _{\mathcal {D} }
	\end{array} \right] \nonumber
\end{equation}
and
\begin{equation}
\left<\mathcal{W}^{(i)}_L,  \pi \right> _{\mathcal {D} } \triangleq
\left[ \begin{array}{ccc}
 		\left< \omega^{(i)}_1, \pi \right> _{\mathcal {D} }~
 		\cdots~
 		\left< \omega^{(i)}_L, \pi \right> _{\mathcal {D} }
	\end{array} \right]^T. \nonumber
\end{equation}
and thus $ \theta^{(i+1)}  $ can be obtained with
\begin{equation} \label{eq_3.37}
\theta^{(i+1)} = \left< \mathcal{W}^{(i)}_L, \overline{\rho}^{(i)} \right> _{\mathcal {D} } ^{-1}
\left<\mathcal{W}^{(i)}_L,  \pi \right> _{\mathcal {D} }.
\end{equation}
Note that the computations of  $ \left< \mathcal{W}^{(i)}_L, \overline{\rho}^{(i)} \right> _{\mathcal {D} } $ and $ \left<\mathcal{W}^{(i)}_L,  \pi \right> _{\mathcal {D} } $ involve many numerical integrals on domain $ \mathcal {D} $, which are computationally expensive. Thus, the Monte-Carlo integration method \cite{peter1978new} is introduced, which is especially competitive on multi-dimensional domain. We now illustrate the Monte-Carlo integration for computing $ \left< \mathcal{W}^{(i)}_L(x, \mu, x') , \overline{\rho}^{(i)}(x, \mu, x') \right> _{\mathcal {D} } $. Let  $ I_\mathcal {D} \triangleq \int_\mathcal {D} d(x, \mu, x')  $, and $ \mathcal {S}_M \triangleq \lbrace (x_{[k]}, \mu_{[k]}, x'_{[k]}) \vert (x_{[k]}, \mu_{[k]}, x'_{[k]}) \in \mathcal {D}, k = 1,2,...,M \rbrace  $   be the set that sampled on domain  $ \mathcal {D} $, where $ M $ is size of sample set $ \mathcal {S}_M $. Then, $ \left< \mathcal{W}^{(i)}_L(x, \mu, x') , \overline{\rho}^{(i)}(x, \mu, x') \right> _{\mathcal {D} } $  is approximately computed with
\begin{flalign} \label{eq_3.38}
& \left< \mathcal{W}^{(i)}_L(x, \mu, x'), \overline{\rho}^{(i)}(x, \mu, x') \right> _{\mathcal {D} }\nonumber \\
& \quad = \int _\mathcal {D} \mathcal{W}^{(i)}_L(x, \mu, x') \overline{\rho}^{(i)}(x, \mu, x') d(x, \mu, x')\nonumber \\
& \quad = \frac{I_\mathcal {D}}{M} \sum _{k=1}^M \mathcal{W}^{(i)}_L(x_{[k]}, \mu_{[k]}, x'_{[k]}) \overline{\rho}^{(i)}(x_{[k]}, \mu_{[k]}, x'_{[k]}) \nonumber \\
& \quad = \frac{I_\mathcal {D}}{M} (W^{(i)})^T Z^{(i)}
\end{flalign}
where
\begin{flalign}
W^{(i)} =& [ \mathcal{W}^{(i)}_L(x_{[1]}, \mu_{[1]}, x'_{[1]})~\cdots~\mathcal{W}^{(i)}_L(x_{[M]}, \mu_{[M]}, x'_{[M]}) ]^T \nonumber \\
Z^{(i)} =& [ ( \overline{\rho}^{(i)}(x_{[1]}, \mu_{[1]}, x'_{[1]}) )^T~\cdots ~( \overline{\rho}^{(i)}(x_{[M]}, \mu_{[M]}, x'_{[M]}) )^T ]^T.\nonumber
\end{flalign}
Similarly,
\begin{flalign} \label{eq_3.39}
& \left<\mathcal{W}^{(i)}_L(x, \mu, x'),  \pi(x,\mu) \right> _\mathcal {D} \nonumber \\
& \quad= \frac{I_\mathcal {D}}{M} \sum _{k=1}^M \left( \mathcal{W}^{(i)}_L(x_{[k]}, \mu_{[k]}, x'_{[k]}) \right)^T \pi(x_{[k]},\mu_{[k]}\nonumber \\
& \quad = \frac{I_\mathcal {D}}{M} (W^{(i)})^T \eta
\end{flalign}
where $ \eta\triangleq \left[ \pi(x_{[1]},\mu_{[1]})~...~\pi(x_{[M]},\mu_{[M]}) \right]^T $. Then, the substitution of \eqref{eq_3.38} and \eqref{eq_3.39} into \eqref{eq_3.37} yields,
\begin{equation} \label{eq_3.40}
\theta^{(i)} = \left[ (W^{(i)})^T Z^{(i)} \right]^{-1} (W^{(i)})^T \eta.
\end{equation}

It is observed that the sample set $ \mathcal {S}_M $  is arbitrary on domain $ \mathcal {D} $, based on which $ W^{(i)} $, $ Z^{(i)} $  and $ \eta $ can be computed and then the unknown parameter vector $ \theta^{(i)} $ is obtained with the expression \eqref{eq_3.40} accordingly.

The above MWR is for solving an iterative equation \eqref{eq_3.11} in the PIQL algorithm. Based on the expression \eqref{eq_3.40}, the implementation procedure of the PIQL algorithm is given as follows:
\noindent \rule{0.49\textwidth}{2pt}
\begin{algorithm} \label{algorithm_3.2}
\emph{
Implementation of PIQL\\
\noindent \rule{0.49\textwidth}{1pt}
\begin{itemize}
\item [$\blacktriangleright$] \emph{Step 1:}
Collect real system data $ (x_k, \mu_k, x'_k) $ for sample set $ \mathcal {S}_M $, and then compute $ \Psi_L(x_k,\mu_k) $ and $ \eta $;
\item [$\blacktriangleright$] \emph{Step 2:}
Let $ u^{(0)}(x) \in \mathfrak{U}(\mathcal {X}) $, and $ i = 0 $;
\item [$\blacktriangleright$] \emph{Step 3:}
Compute $ W^{(i)} $ and $ Z^{(i)} $, and then update parameter vector $\theta^{(i)} $ with \eqref{eq_3.40};
\item [$\blacktriangleright$] \emph{Step 4:}
Update control policy $ \hat{u}^{(i+1)}(x) $ based on \eqref{eq_3.34} with index $ i+1 $;
\item [$\blacktriangleright$] \emph{Step 5:}
If $ \Vert \theta^{(i)} - \theta^{(i-1)} \Vert \leq \xi $ ($ i\geqslant 1 $, $ \xi $ is a small positive number), stop iteration, else, let $i = i+1 $ and go back to Step 3. $\square $
\end{itemize}
}
\end{algorithm}
\noindent \rule{0.49\textwidth}{2pt}

\section{Value Iteration Based Q-learning} \label{Sec_4}

Generally, RL involves two basic frameworks: policy iteration and value iteration. Section \ref{Sec_3} gives a PIQL method for model-free optimal control design of the system \eqref{eq_2.1}.
In this section, we proposed a value iteration based QL (VIQL) algorithm, which is presented as follows:

\noindent \rule{0.49\textwidth}{2pt}
\begin{algorithm} \label{algorithm_4.1}
\emph{
Value iteration based Q-learning\\
\noindent \rule{0.49\textwidth}{1pt}
\begin{itemize}
\item [$\blacktriangleright$] \emph{Step 1:}
Let $ Q^{(0)}(x, \mu) \geqslant 0$ be an initial Q-function. Let $ i = 1 $;
\item [$\blacktriangleright$] \emph{Step 2:}
(\textbf{Policy improvement}) Update control policy with:
\begin{equation} \label{eq_4.1}
u^{(i)} (x) \triangleq \arg \min_\mu Q^{(i-1)}(x, \mu);
\end{equation}
\item [$\blacktriangleright$] \emph{Step 3:}
(\textbf{Policy evaluation}) Solve the iterative equation
\begin{equation} \label{eq_4.2}
Q^{(i)}(x_t, \mu) \triangleq \int_{t}^{t'}\mathcal {R} (x(\tau),\mu(\tau)) d \tau + Q^{(i-1)}(x'_t, u^{(i)})
\end{equation}
for unknown Q-function $ Q^{(i)} $;
\item [$\blacktriangleright$] \emph{Step 4:}
Let $ i = i+1 $, go back to Step 2 and continue.
$\square $
\end{itemize}
}
\end{algorithm}
\noindent \rule{0.49\textwidth}{2pt}

\begin{theorem}\label{theorem_4.1}
For $ \forall (x, \mu) \in \mathcal{X} \times \mathcal{U} $, let $ \{ Q^{(i)}(x, \mu) \} $ be the sequence generated by Algorithm \ref{algorithm_4.1}. Let $ Q^{(0)}(x, \mu) $ be an arbitrary Q-function of a stable control policy, then\\
1) $ Q^{(i)}(x,\mu) \geqslant Q^{(i+1)}(x,\mu) \geqslant Q^*(x,\mu) $ \\
2) $ Q^{(i)}(x,\mu) \rightarrow Q^*(x,\mu) $ and $ u^{(i)}(x) \rightarrow u^*(x) $ as $ i \rightarrow \infty $.
\end{theorem}
\noindent \textbf{Proof.} 1) The proof is by mathematical induction. 

Without loss of generality, let $ Q^{(0)}(x, \mu) $ denotes the Q-function of a stable control policy $ \upsilon(x) $. Then, it follows from the definition of Q-function \eqref{eq_3.6} and \eqref{eq_3.7a} that
 \begin{flalign} \label{eq_4.3}
Q^{(0)}(x_t, \mu) = & Q_\upsilon(x_t, \mu) \nonumber \\
= & \int_{t}^{t'}\mathcal {R} (x(\tau),\mu(\tau)) d \tau + Q^{(0)}(x_t', \upsilon).
\end{flalign}
According to \eqref{eq_4.1}-\eqref{eq_4.3}, 
 \begin{flalign}
Q^{(1)}(x_t, \mu) = & \int_{t}^{t'}\mathcal {R} (x(\tau),\mu(\tau)) d \tau + Q^{(0)}(x_t', u^{(1)}) \nonumber \\
= & \int_{t}^{t'}\mathcal {R} (x(\tau),\mu(\tau)) d \tau + \min_\mu Q^{(0)}(x'_t, \mu) \nonumber \\
\leqslant & \int_{t}^{t'}\mathcal {R} (x(\tau),\mu(\tau)) d \tau + Q^{(0)}(x'_t, \upsilon) \nonumber \\
= & Q^{(0)}(x_t, \mu) \nonumber
\end{flalign}
which means that the inequality $ Q^{(i)}(x,\mu) \geqslant Q^{(i+1)}(x,\mu) $ holds for $ i = 0 $.

Assume that the inequality $ Q^{(i)}(x,\mu) \geqslant Q^{(i+1)}(x,\mu) $ holds for $ i = l-1 $, i.e.,
\begin{equation} \label{eq_4.4}
Q^{(l)}(x, \mu) \leqslant Q^{(l-1)}(x, \mu).
\end{equation}
Based on the expression \eqref{eq_4.1},
\begin{equation} \label{eq_4.5}
Q^{(l)}(x, u^{(l)}) = \min_\mu Q^{(l)}(x, \mu) \leqslant Q^{(l)}(x, \mu)
\end{equation}
for $ \forall x, \mu $.
Then, it follows from \eqref{eq_4.2}, \eqref{eq_4.4} and \eqref{eq_4.5} that
\begin{flalign} \label{eq_4.6}
Q^{(l+1)}(x_t, \mu) & = \int_{t}^{t'}\mathcal {R} (x(\tau),\mu(\tau)) d \tau + Q^{(l)}(x'_t, u^{(l+1)}) \nonumber \\
& = \int_{t}^{t'}\mathcal {R} (x(\tau),\mu(\tau)) d \tau + \min_\mu Q^{(l)}(x'_t, \mu) \nonumber \\
& \leqslant \int_{t}^{t'}\mathcal {R} (x(\tau),\mu(\tau)) d \tau + Q^{(l)}(x'_t, u^{(l)}) \nonumber \\
& \leqslant \int_{t}^{t'}\mathcal {R} (x(\tau),\mu(\tau)) d \tau + Q^{(l-1)}(x'_t, u^{(l)}) \nonumber \\
& = Q^{(l)}(x_t, \mu)
\end{flalign}
which means that the inequality $ Q^{(i)}(x,\mu) \geqslant Q^{(i+1)}(x,\mu) $ holds for $ i = l $.

Next, we will prove $ Q^{(i)}(x,\mu) \geqslant Q^*(x,\mu) $. Considering $ Q^{(i-1)}(x,\mu) \geqslant Q^{(i)}(x,\mu) $ holds for $ \forall (x,\mu) \in \mathcal{X} \times \mathcal{U} $, then
\begin{equation} \label{eq_4.7}
Q^{(i-1)}(x'_t,u^{(i)}) \geqslant Q^{(i)}(x'_t,u^{(i)})
\end{equation}
Combining \eqref{eq_4.2} and \eqref{eq_4.7}, there is
\begin{flalign} \label{eq_4.10}
Q^{(i)}(x_t, \mu) = & \int_{t}^{t'}\mathcal {R} (x(\tau),\mu(\tau)) d \tau + Q^{(i-1)}(x'_t, u^{(i)})\nonumber \\
\geqslant & \int_{t}^{t'}\mathcal {R} (x(\tau),\mu(\tau)) d \tau + Q^{(i)}(x'_t, u^{(i)})
\nonumber \\
= & \int_{t}^{t'}\mathcal {R} (x(\tau),\mu(\tau)) d \tau + V^{(i)}(x'_t) \nonumber \\
\geqslant & \int_{t}^{t'}\mathcal {R} (x(\tau),\mu(\tau)) d \tau + V^*(x'_t) \nonumber \\
= & Q^*(x_t, \mu).
\end{flalign}
The part 1) of Theorem \ref{theorem_4.1} is proved.

2) The part 2) of Theorem \ref{theorem_4.1} can be proved similarly as that for the part 2) of Theorem \ref{theorem_3.3}, and it is omitted for briefness.
The proof is completed.
$ \square $

It is noted that an iterative equation \eqref{eq_4.2} should be solved in each iteration of the VIQL algorithm. Similar with Subsection \ref{Sec_3.3}, the MWR can be used to solve the equation \eqref{eq_4.2}. To avoid repeat, we omit the detailed derivation of the MWR and give the parameter update law directly as
\begin{equation} \label{eq_4.11}
\theta^{(i)} = [ W^T Z ]^{-1} W^T (\eta + Z^{(i)} \theta^{(i-1)})
\end{equation}
where the notations are given by
\begin{flalign}
W &= [ \mathcal{W}_L(x_{[1]}, \mu_{[1]}) ~\cdots~\mathcal{W}_L(x_{[M]}, \mu_{[M]}) ]^T \nonumber \\
Z &= [ \overline{\rho}^T(x_{[1]}, \mu_{[1]})~\cdots ~ \overline{\rho}^T(x_{[M]}, \mu_{[M]}) ]^T \nonumber \\
Z^{(i)} &= [ \overline{\rho}^T(x'_{[1]}, \hat{u}^{(i)}(x'_{[1]}))~\cdots~ \overline{\rho}^T(x'_{[M]}, \hat{u}^{(i)}(x'_{[M]})) ]^T \nonumber \\
\eta &= \left[ \pi(x_{[1]},\mu_{[1]})~...~\pi(x_{[M]},\mu_{[M]}) \right]^T \nonumber
\end{flalign}
with $ \mathcal{W}_L(x, \mu) =  [\omega_1(x, \mu)~~\cdots~~\omega_L(x, \mu)]^T$ be the weighted function vector,
$ \overline{\rho}(x, \mu) = [\overline{\rho}_1(x, \mu)~\cdots~\overline{\rho}_L(x, \mu)] $ with $ \overline{\rho}_j(x, \mu) =  \psi_j(x,\mu), j = 1,...,L $, and $ \pi(x_t, \mu) = \int_{t}^{t'}\mathcal {R} (x(\tau),\mu(\tau)) d \tau $.

Based on the parameter update law \eqref{eq_4.11}, the implementation procedure of the VIQL algorithm is presented below:

\noindent \rule{0.49\textwidth}{2pt}
\begin{algorithm} \label{algorithm_4.2}
\emph{
Implementation of VIQL\\
\noindent \rule{0.49\textwidth}{1pt}
\begin{itemize}
\item [$\blacktriangleright$] \emph{Step 1:}
Collect real system data $ (x_k, \mu_k, x'_k) $ for sample set $ \mathcal {S}_M $, and then compute $ W, Z $ and $ \eta $;
\item [$\blacktriangleright$] \emph{Step 2:}
Let $\theta^{(0)} $ be the initial parameter vector, and $ i = 1 $;
\item [$\blacktriangleright$] \emph{Step 3:}
Update control policy $ \hat{u}^{(i)}(x) $ with \eqref{eq_3.34};
\item [$\blacktriangleright$] \emph{Step 4:}
Compute $ Z^{(i)} $, and then update parameter vector $\theta^{(i)} $ with \eqref{eq_4.11};
\item [$\blacktriangleright$] \emph{Step 5:}
If $ \Vert \theta^{(i)} - \theta^{(i-1)} \Vert \leq \xi $ ($ \xi $ is a small positive number), stop iteration, else, let $i = i+1 $ and go back to Step 3. $\square $
\end{itemize}
}
\end{algorithm}
\noindent \rule{0.49\textwidth}{2pt}

\begin{remark} \label{remark3}
\emph{
There are several similarities and differences between PIQL and VIQL algorithms, which are summarized as follows: 1) Both of them generate a non-increasing Q-function sequence, which converges to the optimal  Q-function.
2) Both algorithms are model-free method, which learns the optimal control policy from real system data rather than using mathematical model of system \eqref{eq_2.1}.
3) Both of them are off-policy RL method, where the system data can be generated by arbitrary behavior control policies.
4) Their implementations are offline learning procedure, and then the convergent control will be employed for real-time control.
5) PIQL algorithm requires an initial admissible control policy, while it is not a necessity for VIQL algorithm.
6) For RL methods, the policy iteration has a quadratic convergence rate, while value iteration has a linear convergence rate. This implies that PIQL algorithm converges much faster than VIQL algorithm.
$ \square $ }
\end{remark}

\section{Q-learning for LQR Problem} \label{Sec_5}
In the section, we simplify the developed PIQL and VIQL methods for solving the model-free linear quadratic regulation (LQR) problem. Consider a linear version of system \eqref{eq_2.1}:
\begin{equation}\label{eq_5.1}
\dot{x}(t) = Ax(t) + Bu(t), x(0) = x_0
\end{equation}
where $ A \in \mathbb{R}^{n \times n} $ and $ B \in \mathbb{R}^{n \times m} $ are unknown matrices,
and the quadratic cost function:
\begin{equation} \label{eq_5.2}
J(x_0,u) = \int_{0}^{\infty} {\left( \Vert x(t) \Vert _S^2 + \Vert u(t) \Vert _W^2 \right)} dt
\end{equation}
where matrices $ S,W>0 $.

For the LQR problem of system \eqref{eq_5.1} with cost function \eqref{eq_5.2}, its optimal Q-function can be given by
\begin{equation}\label{eq_5.3}
Q^*(x,\mu) =
\left[ \begin{array} {*{3}{>{\displaystyle}c}}
 		x\\
 		\mu
	\end{array} \right]^T
 G
 \left[ \begin{array} {*{3}{>{\displaystyle}c}}
 		x\\
 		\mu
	\end{array} \right]
\end{equation}
where  $ G \geqslant 0 $ is a block matrix denoted as
\begin{equation}
G \triangleq
\left[ \begin{array} {*{3}{>{\displaystyle}cc}}
 		G_{11} & G_{12}\\
 		G_{21} & G_{22}
	\end{array} \right].
\nonumber
\end{equation}
According to \eqref{eq_3.9}, its optimal control policy is given by
\begin{equation} \label{eq_5.4}
u^*(x) =\arg\min_\mu Q^*(x, \mu) = -G^{-1}_{22}G_{12}^Tx.
\end{equation}
Thus, denote the Q-function $ Q^{(i)}(x, \mu) $ as
\begin{equation}\label{eq_5.5}
Q^{(i)}(x,\mu) =
\left[ \begin{array} {*{3}{>{\displaystyle}c}}
 		x\\
 		\mu
	\end{array} \right]^T
 G^{(i)}
 \left[ \begin{array} {*{3}{>{\displaystyle}c}}
 		x\\
 		\mu
	\end{array} \right]
\end{equation}
where  $ G^{(i)} $ is given by
\begin{equation}
G^{(i)} \triangleq
\left[ \begin{array} {*{3}{>{\displaystyle}cc}}
 		G^{(i)}_{11} & G^{(i)}_{12}\\
 		G^{(i)}_{21} & G^{(i)}_{22}
	\end{array} \right].
\nonumber
\end{equation}

\subsection{PIQL for LQR Problem} \label{Sec_5.1}

By using the expression \eqref{eq_5.5}, the PIQL Algorithm \ref{algorithm_3.1} can be simplified for solving the model-free LQR problem of system \eqref{eq_5.1}, which is given as follows:

\noindent \rule{0.49\textwidth}{2pt}
\begin{algorithm} \label{algorithm_5.1}
\emph{
PIQL for LQR problem\\
\noindent \rule{0.49\textwidth}{1pt}
\begin{itemize}
\item [$\blacktriangleright$] \emph{Step 1:}
Let $ u^{(0)}(x) \in \mathfrak{U}(\mathcal {X}) $ be an initial stabilizing control policy, and $ i = 0 $;
\item [$\blacktriangleright$] \emph{Step 2:}
(\textbf{Policy evaluation}) Solve the equation
\begin{flalign} \label{eq_5.6}
& \left[ \begin{array} {*{3}{>{\displaystyle}c}}
 		x_t\\
 		\mu
	\end{array} \right]^T
 G^{(i)}
 \left[ \begin{array} {*{3}{>{\displaystyle}c}}
 		x_t\\
 		\mu
	\end{array} \right]
	=
	\int_{t}^{t'}\mathcal {R} (x(\tau),\mu(\tau)) d \tau \nonumber \\
	& \qquad +
	\left[ \begin{array} {*{3}{>{\displaystyle}c}}
 		x'_t\\
 		u^{(i)}
	\end{array} \right]^T
 G^{(i)}
 \left[ \begin{array} {*{3}{>{\displaystyle}c}}
 		x'_t\\
 		u^{(i)}
	\end{array} \right]
\end{flalign}
for unknown unknown matrix $ G^{(i)} $;
\item [$\blacktriangleright$] \emph{Step 3:}
(\textbf{Policy improvement}) Update control policy with
\begin{equation} \label{eq_5.7}
u^{(i+1)}(x) = -[G_{22}^{(i)}]^{-1}[G_{12}^{(i)}]^Tx;
\end{equation}
\item [$\blacktriangleright$] \emph{Step 4:}
Let $ i = i+1 $, go back to Step 2 and continue.
$\square $
\end{itemize}
}
\end{algorithm}
\noindent \rule{0.49\textwidth}{2pt}

According to Theorem \ref{theorem_3.3}, Algorithm \ref{algorithm_5.1} generates a non-increasing matrix sequence $ \{ G^{(i)} \} $ that converges to $ G $.

For each iteration of Algorithm \ref{algorithm_5.1}, the iterative equation \eqref{eq_5.6} should be solved for the unknown unknown matrix $ G^{(i)} $. It is observed that $ G^{(i)} $ is a $ (n+m) \times (n+m) $ symmetric matrix, which has $ (n+m)(n+m+1)/2 $ unknown parameters. Letting
\begin{eqnarray} \label{eq_5.8}
\Psi_L(x, \mu) = [x_1^2~~x_1x_2~~...~~x_1\mu_1~~...~~x_1\mu_m~~x_2^2\nonumber \\
~~x_2\mu_1~~...~~\mu_m^2]^T
\end{eqnarray}
and
\begin{eqnarray} \label{eq_5.9}
\theta^{(i)} = [(g_{1,1}^{(i)})^2 ~~2g_{1,2}^{(i)}~~...~~2g_{1,n+1}^{(i)}~~...~~2g_{1,n+m}^{(i)}~~(g_{2,2}^{(i)})^2 \nonumber \\
~~2g_{2,n+m}^{(i)}~~...~~(g_{n+m,n+m}^{(i)})^2]^T
\end{eqnarray}
the iterative equation \eqref{eq_5.6} can be equivalently rewritten as
\begin{equation} \label{eq_5.10}
[\Psi_L(x_t, \mu) - \Psi_L(x'_t, \hat{u}^{(i)}(x'_t))]^T \theta^{(i)} = \pi(x_t, \mu)
\end{equation}
where $ \pi(x_t, \mu) \triangleq \int_{t}^{t'}\mathcal {R} (x(\tau),\mu(\tau)) d \tau $. By using the data set $ \mathcal {S}_M $ collected from real system, the following least-square scheme is obtained for updating the unknown parameter vector $ \theta^{(i)} $
\begin{equation} \label{eq_5.11}
\theta^{(i)} = \left[ (Z^{(i)})^T Z^{(i)} \right]^{-1} (Z^{(i)})^T \eta
\end{equation}
where
\begin{flalign}
Z^{(i)} =& [ \Psi_L(x_{[1]}, \mu_{[1]}) - \Psi_L(x'_{[1]}, \hat{u}^{(i)}(x'_{[1]}))~\nonumber \\
&...~\Psi_L(x_{[M]}, \mu_{[M]}) -  \Psi_L(x'_{[M]}, \hat{u}^{(i)}(x'_{[M]})) ]^T \nonumber \\
\eta =& [ \pi(x_{[1]},\mu_{[1]})~...~\pi(x_{[M]},\mu_{[M]}) ]^T. \nonumber
\end{flalign}
With the least-square scheme \eqref{eq_5.11}, the implementation procedure of Algorithm \ref{algorithm_5.1} will be obtained similarly with Algorithm \ref{algorithm_3.2}.
\begin{remark} \label{remark4}
\emph{
For the LQR problem of the linear system \eqref{eq_5.1}, there is no residual error in the expression \eqref{eq_5.10} because the Q-function $ Q(x,\mu) $ can be exactly represented by basic function vector \eqref{eq_5.9}.
$ \square $ }
\end{remark}

\subsection{VIQL for LQR Problem} \label{Sec_5.2}
With the expression \eqref{eq_5.5}, the VIQL Algorithm \ref{algorithm_4.1} can be simplified for solving the model-free LQR problem of system \eqref{eq_5.1}, which is given as follows:

\noindent \rule{0.49\textwidth}{2pt}
\begin{algorithm} \label{algorithm_5.2}
\emph{
VIQL for LQR problem\\
\noindent \rule{0.49\textwidth}{1pt}
\begin{itemize}
\item [$\blacktriangleright$] \emph{Step 1:}
Let $ G^{(0)} \geqslant 0 $ and $ i = 1 $;
\item [$\blacktriangleright$] \emph{Step 2:}
(\textbf{Policy improvement}) Update control policy with:
\begin{equation} \label{eq_5.12}
u^{(i)}(x) = -[G_{22}^{(i-1)}]^{-1}[G_{12}^{(i-1)}]^Tx;
\end{equation}
\item [$\blacktriangleright$] \emph{Step 3:}
(\textbf{Policy evaluation}) Solve the iterative equation
\begin{flalign} \label{eq_5.13}
& \left[ \begin{array} {*{3}{>{\displaystyle}c}}
 		x_t\\
 		\mu
	\end{array} \right]^T
 G^{(i)}
 \left[ \begin{array} {*{3}{>{\displaystyle}c}}
 		x_t\\
 		\mu
	\end{array} \right]
	=
	\int_{t}^{t'}\mathcal {R} (x(\tau),\mu(\tau)) d \tau \nonumber \\
	& \qquad +
	\left[ \begin{array} {*{3}{>{\displaystyle}c}}
 		x'_t\\
 		u^{(i)}
	\end{array} \right]^T
 G^{(i-1)}
 \left[ \begin{array} {*{3}{>{\displaystyle}c}}
 		x'_t\\
 		u^{(i)}
	\end{array} \right]
\end{flalign}
for unknown unknown matrix $ G^{(i)} $;
\item [$\blacktriangleright$] \emph{Step 4:}
Let $ i = i+1 $, go back to Step 2 and continue.
$\square $
\end{itemize}
}
\end{algorithm}
\noindent \rule{0.49\textwidth}{2pt}

Based on Theorem \ref{theorem_4.1}, Algorithm \ref{algorithm_5.2} generates a non-increasing matrix sequence $ \{ G^{(i)} \} $ that converges to $ G $. By using the expressions \eqref{eq_5.8} and \eqref{eq_5.9}
the iterative equation \eqref{eq_5.13} can be equivalently rewritten as
\begin{equation} \label{eq_5.14}
\Psi^T_N(x_t, \mu) \theta^{(i)} = \pi(x_t, \mu) + \Psi^T_N(x'_t, \hat{u}^{(i)}(x'_t)) \theta^{(i-1)}
\end{equation}
where $ \pi(x_t, \mu) = \int_{t}^{t'}\mathcal {R} (x(\tau),\mu(\tau)) d \tau $.
By using the data set $ \mathcal {S}_M $ collected from real system, the following least-square scheme is obtained for the updating unknown parameter vector $ \theta^{(i)} $
\begin{equation} \label{eq_5.15}
\theta^{(i)} = [ Z^T Z ]^{-1} Z^T (\eta + Z^{(i)} \theta^{(i-1)})
\end{equation}
where  the notations are given by
\begin{flalign}
Z =& [\Psi_L(x_{[1]}, \mu_{[1]})~\cdots~ \Psi_L(x_{[M]}, \mu_{[M]})]^T\nonumber \\
Z^{(i)} =& [ \Psi_L(x'_{[1]}, \hat{u}^{(i)}(x'_{[1]}))~\cdots~ \Psi_L(x'_{[M]}, \hat{u}^{(i)}(x'_{[M]})) ]^T \nonumber \\
\eta=& \left[ \pi(x_{[1]},\mu_{[1]})~...~\pi(x_{[M]},\mu_{[M]}) \right]^T. \nonumber
\end{flalign}

With the least-square scheme \eqref{eq_5.15}, the implementation procedure of Algorithm \ref{algorithm_5.2} will be obtained similarly with Algorithm \ref{algorithm_4.2}, which is omitted for briefness.

\section{Simulation Studies} \label{Sec_6}
To test the effectiveness of the proposed QL algorithms, computer simulation studies are conducted on a linear F-16 aircraft plant and a numerical nonlinear system. For all simulations, the algorithm stop accuracy is set as  $ \xi  = 10^{-5}$, the system data set $ \mathcal {S}_M $ is collected randomly and its size is set as $ M = 100 $.

\subsection{Example 1: Linear F-16 Aircraft Plant} \label{Sec_6.1}
Consider a linear F16 aircraft plant \cite{stevens2003aircraft, vamvoudakis2014online, vamvoudakis2010online}, where the system dynamics is described with \eqref{eq_5.1}, and the system matrices given by:
\begin{flalign}
A =\left[ \begin{array}{ccc}
 		-1.01887 & 0.90506 & -0.00215 \\
 		0.82225 & -1.07741 & -0.17555 \\
 		0 & 0 & -1
	\end{array} \right],
B =
\left[ \begin{array}{ccc}
 		0 \\
 		0 \\
 		1
	\end{array} \right] \nonumber
\end{flalign}
where the system state vector is $ x = [\alpha~q~\delta_e  ]^T $, $ \alpha $ denotes the angle of attack, $ q $ is the pitch rate, $ \delta_e $ is the elevator deflection angle, and the control input $ u $ is the elevator actuator voltage. Select $ S, W $ as unit matrices for quadratic cost function \eqref{eq_5.2}.
Then, solve the associated Riccati algebraic equation with the MATLAB command CARE, we can obtain the optimal control policy given by $ u^*(x) = Kx $, where $ K = [-0.1352~~-0.1501~~0.4329] $ is the optimal control gain.

For the LQR problem, it follows from \eqref{eq_5.8} and \eqref{eq_5.9} that the the basic function set is
\begin{eqnarray} \label{eq_6.1}
\Psi_L(x, \mu) = [x_1^2~~x_1x_2~~x_1x_3~~x_1\mu~~x_2^2~~x_2x_3 \nonumber \\
x_2\mu ~~ x_3^2~~ x_3\mu ~~ \mu^2]^T
\end{eqnarray}
and the parameter vector is
\begin{eqnarray}
\theta^{(i)} = [(g_{1,1}^{(i)})^2 ~~2g_{1,2}^{(i)}~~2g_{1,3}^{(i)}~~2g_{1,4}^{(i)}~~(g_{2,2}^{(i)})^2 ~~2g_{2,3}^{(i)} \nonumber \\
2g_{2,4}^{(i)}~~(g_{3,3}^{(i)})^2 ~~2g_{3,4}^{(i)}~~(g_{4,4}^{(i)})^2]^T. \nonumber
\end{eqnarray}
According to the policy improvement rule \eqref{eq_5.7}, the iterative control gain is $ K^{(i)} = [k_1^{(i)}~~k_2^{(i)}~~k_3^{(i)}] $ $ = - (g_{4,4}^{(i)})^{-1}[g_{1,4}^{(i)}~~g_{2,4}^{(i)}~~g_{3,4}^{(i)}]$ $ = - 0.5 (\theta_{10}^{(i)})^{-1} [\theta_4^{(i)}~~\theta_7^{(i)}~~\theta_9^{(i)}] $.
\begin{figure}
\centering	\includegraphics[width=3.2in]{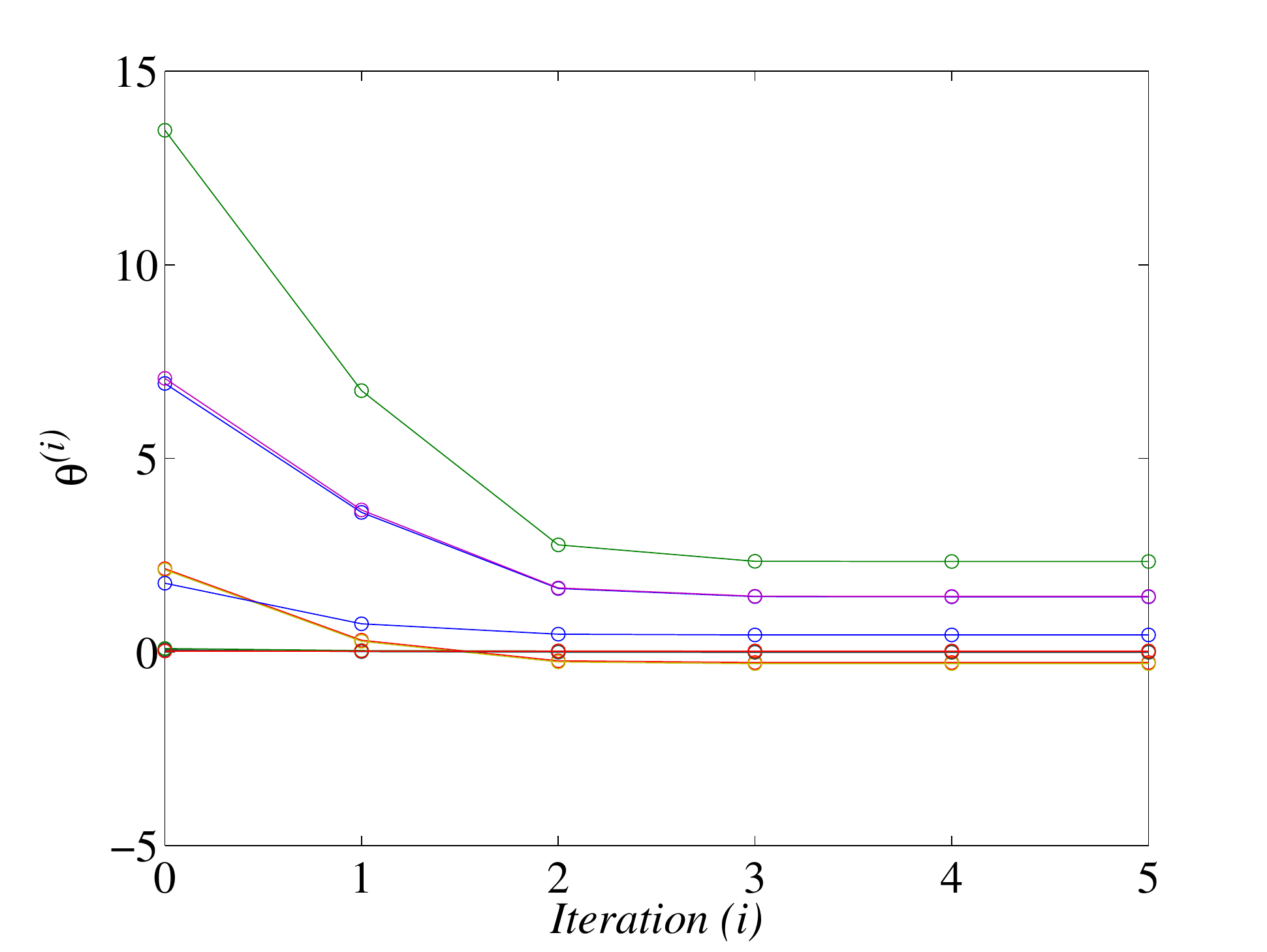}
	\caption{For example 1, all parameters of vector $ \theta^{(i)} $ obtained by PIQL algorithm.}
		\label{fig1}
\end{figure}
\begin{figure}
\centering	\includegraphics[width=3.2in]{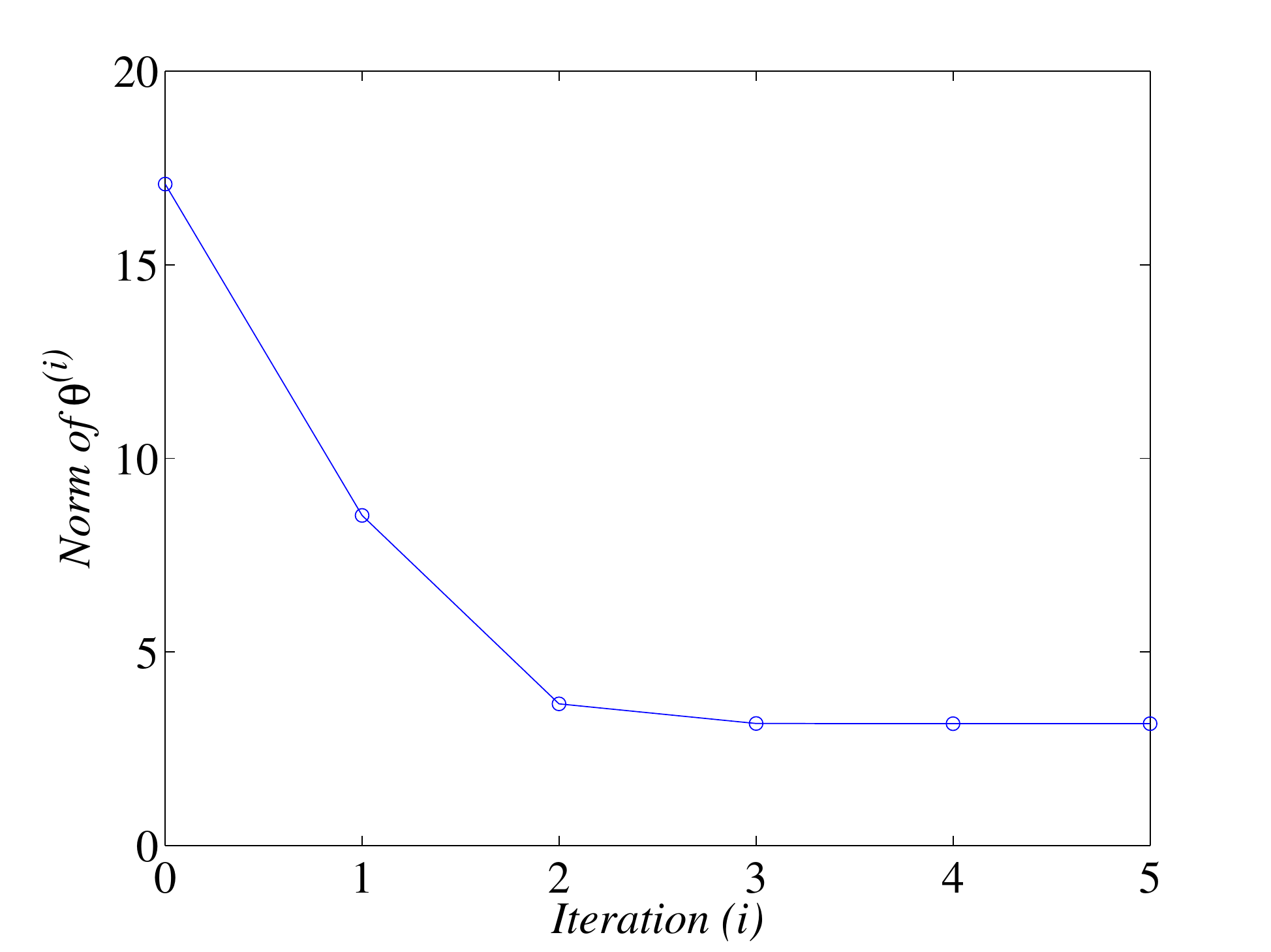}
	\caption{For example 1, the norm $ \Vert \theta^{(i)} \Vert $ obtained by PIQL algorithm.}
		\label{fig2}
\end{figure}
\begin{figure}
\centering	\includegraphics[width=3.2in]{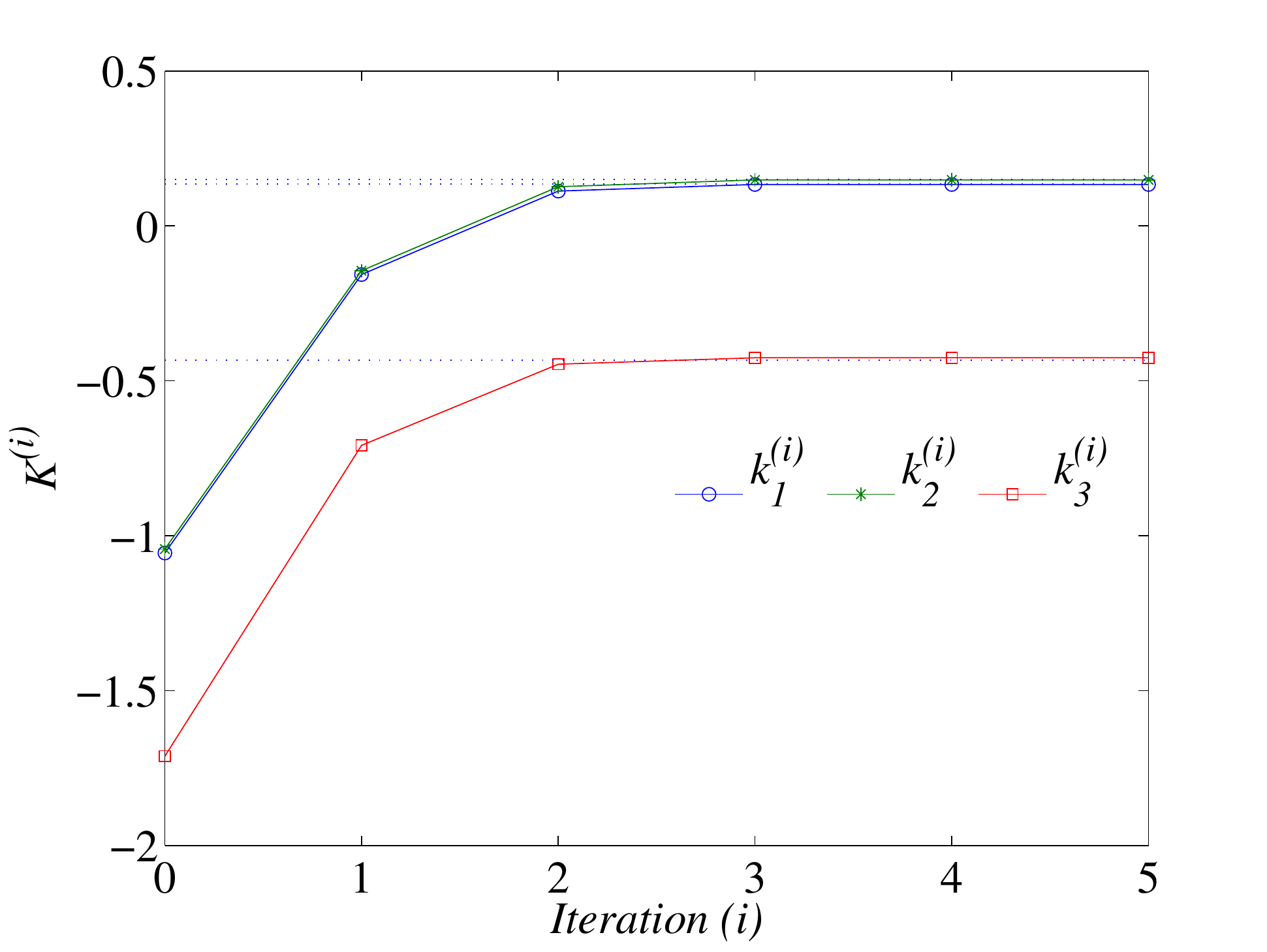}
	\caption{For example 1, the iterative control gain $ K^{(i)} $ obtained by PIQL algorithm.}
		\label{fig3}
\end{figure}

First, the PIQL algorithm (i.e., Algorithm \ref{algorithm_3.2}) with parameters update law \eqref{eq_5.11} is used to solve this model-free LQR problem. Figures \ref{fig1}-\ref{fig3} gives the simulation results, where the PIQL algorithm achieves convergence at $ i = 5 $ iteration. Figures \ref{fig1} and \ref{fig2} show the parameter vector $ \theta^{(i)} $ and its norm $ \Vert \theta^{(i)} \Vert $ respectively, where the parameter vector converges to
\begin{eqnarray}
\theta^{(5)} = [1.4250~~
    2.3374~~
   -0.2734~~
   -0.0068~~
    1.4355~~\nonumber \\
   -0.3034~~
   -0.0076~~
    0.4375~~
    0.0216~~
    0.0254]^T.\nonumber
\end{eqnarray}
Figure \ref{fig3} shows the control gain $ K^{(i)} $ at each iteration, where the dot lines represent idea value of the optimal control gain $ K $. It is observed that $ K^{(i)} $ converges to
\begin{eqnarray}
K^{(5)} = [0.1343~~0.1488~~-0.4256]\nonumber
\end{eqnarray}
which approaches to the optimal control gain $ K $.
\begin{figure}
\centering	\includegraphics[width=3.2in]{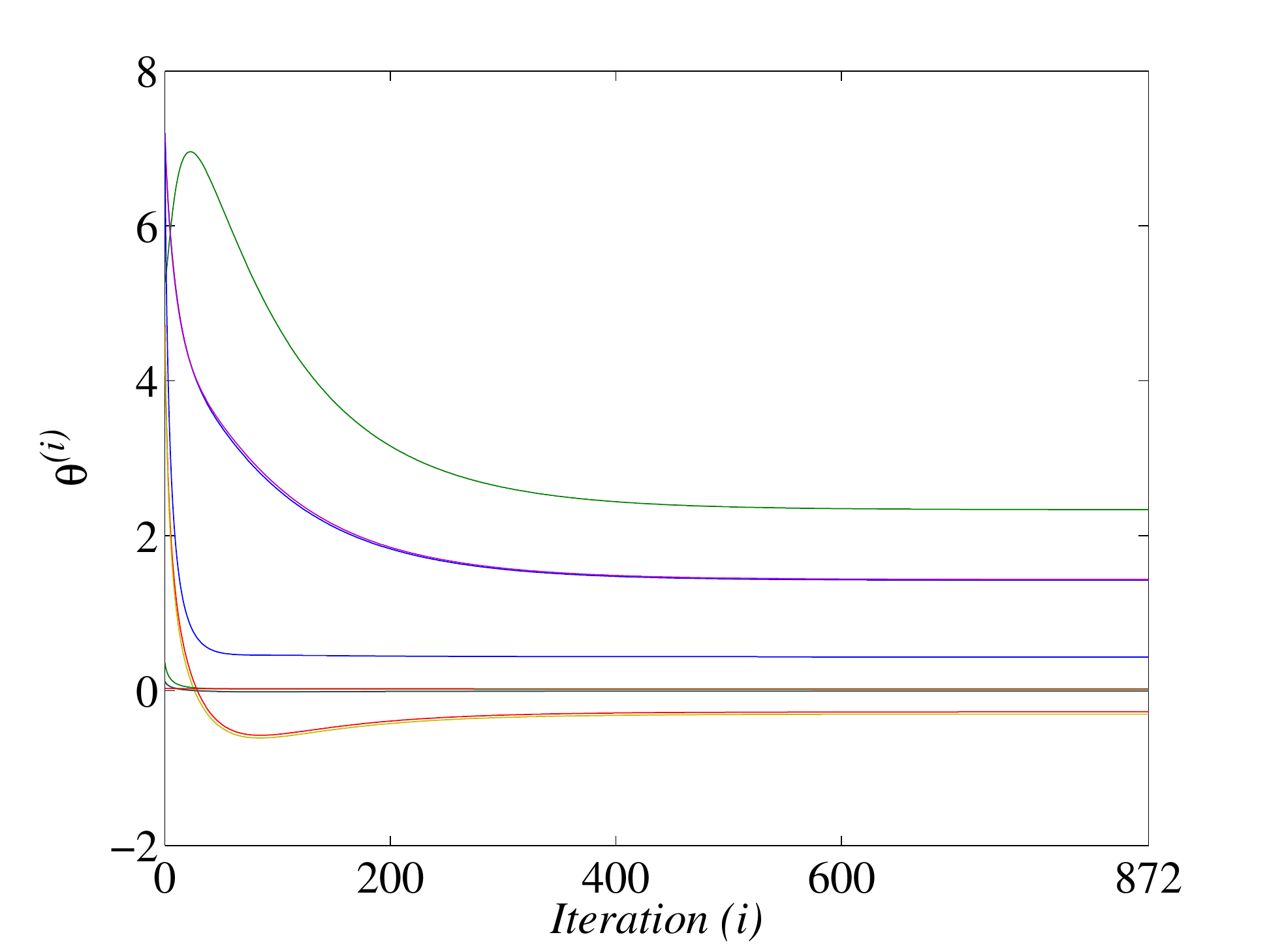}
	\caption{For example 1, all parameters of vector $ \theta^{(i)} $ obtained by VIQL algorithm.}
		\label{fig4}
\end{figure}
\begin{figure}
\centering	\includegraphics[width=3.2in]{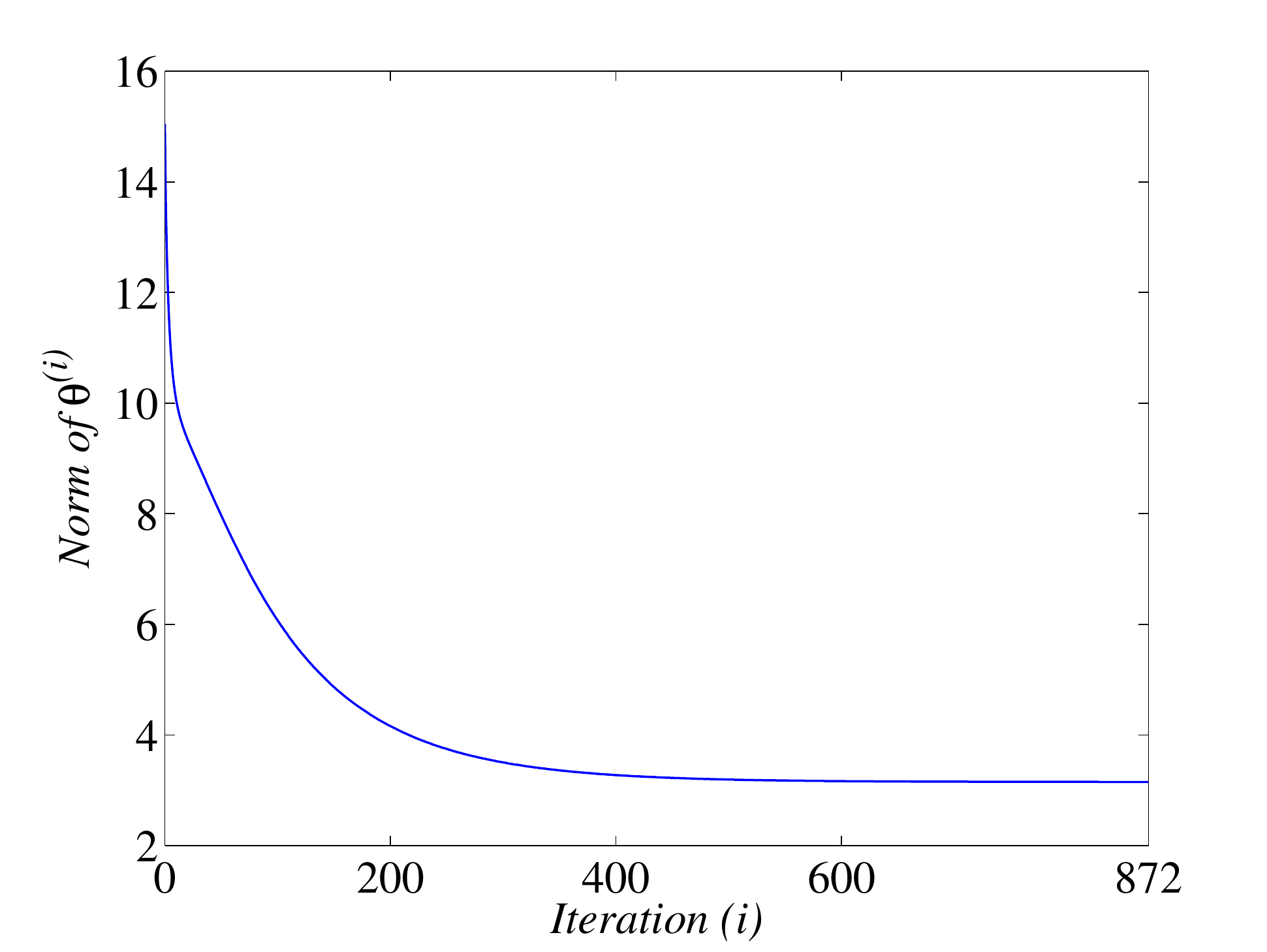}
	\caption{For example 1, the norm $ \Vert \theta^{(i)} \Vert $ obtained by VIQL algorithm.}
		\label{fig5}
\end{figure}
\begin{figure}
\centering	\includegraphics[width=3.2in]{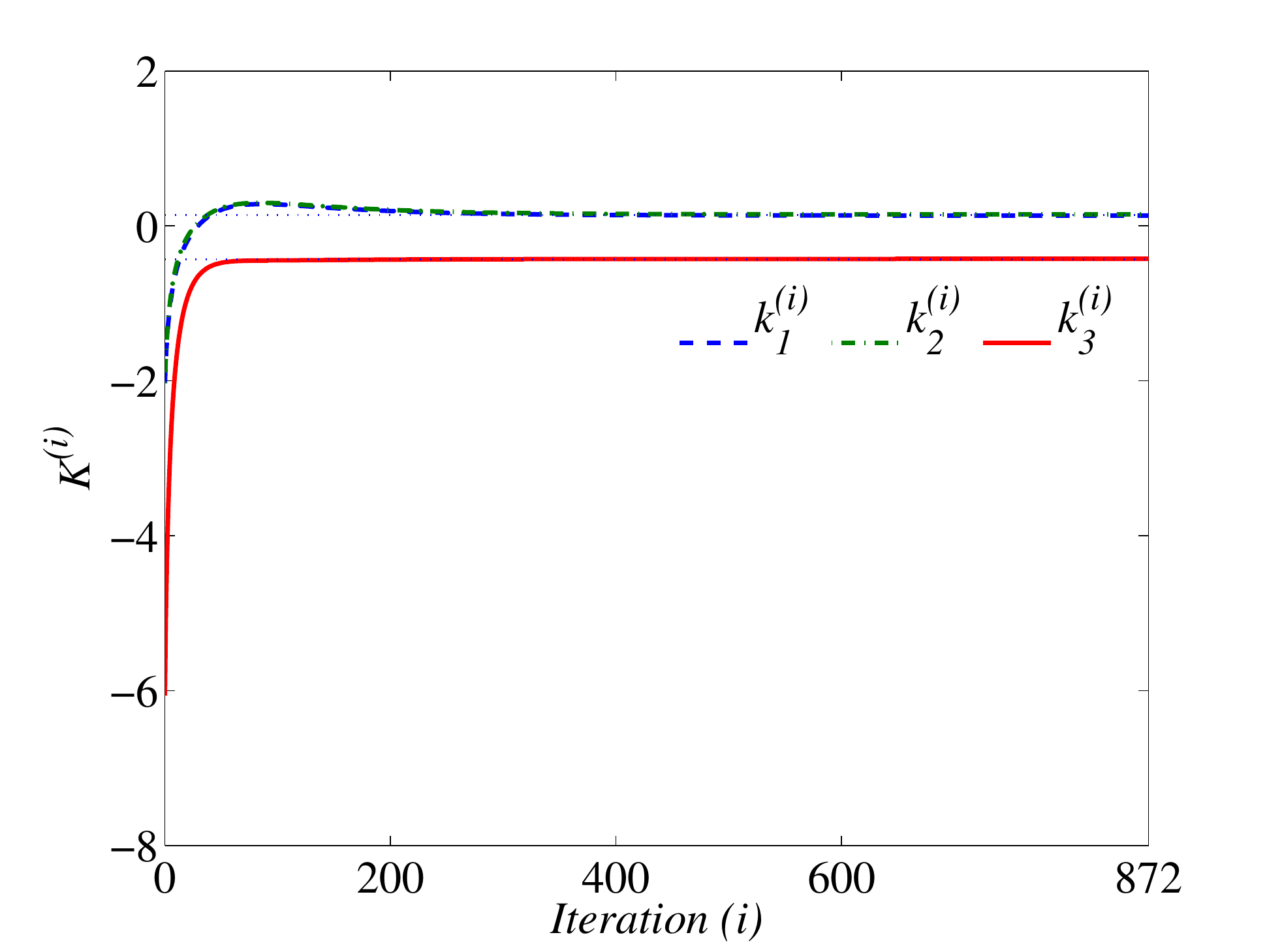}
	\caption{For example 1, the iterative control gain $ K^{(i)} $ obtained by VIQL algorithm.}
		\label{fig6}
\end{figure}

Next, with the same basic function vector \eqref{eq_6.1}, the VIQL algorithm (i.e., Algorithm \ref{algorithm_4.2}) with parameters update law \eqref{eq_5.15} is employed to solve this model-free LQR problem. It is observed that the VIQL algorithm achieves convergence at $ i = 872$ iteration, where the parameter vector $ \theta^{(i)} $ converges to
\begin{eqnarray}
\theta^{(872)} = [
    1.4254~~
    2.3382~~
   -0.2735~~
   -0.0068~~
    1.4359~~\nonumber \\
   -0.3035~~
   -0.0076~~
    0.4375~~
    0.0216~~
    0.0254]^T.\nonumber
\end{eqnarray}
and the iterative control gain $ K^{(i)} $ converges to
\begin{eqnarray}
K^{(872)} = [0.1344~~0.1489~~-0.4256] \nonumber
\end{eqnarray}
which approaches to the optimal control gain $ K $.
Figure \ref{fig4}-\ref{fig5} give the the parameter vector $ \theta^{(i)} $ and the norm $ \Vert \theta^{(i)} \Vert $ at each iteration, respectively. Figure \ref{fig6} shows the control gain $ K^{(i)} $ at each iteration, where the dot lines represent idea value of the optimal control gain $ K $.

From the simulation results, it is found that the PIQL algorithm achieves much faster convergence than VIQL algorithm.

\subsection{Example 2: Numerical Nonlinear System} \label{Sec_6.2}
This numerical example is constructed by using the converse HJB approach \cite{nevistic1996optimality}. The system model is given as follows:
\begin{flalign}
\dot{x} = \left[ \begin{array} {*{3}{>{\displaystyle}c}}
 		-x_1 + x_2\\
 		-0.5(x_1 + x_2) + 0.5x_1^2x_2
	\end{array} \right] +
\left[ \begin{array} {*{3}{>{\displaystyle}c}}
 		0 \\
 		x_1
	\end{array} \right] u \label{eq_6.2}
\end{flalign}
where $ x_0 = [0.1~~0.1]^T $. With the choice of $ Q(x)=x^Tx $  and $ W(u)=u^2 $  in the cost function \eqref{eq_2.2}, the optimal control policy is  $ u^*(x) = -x_1x_2 $.

Select the basic function vector as
\begin{eqnarray} \label{eq_6.3}
\Psi_L(x, \mu) =
[x_1^2~~x_1x_2~~x_2^2~~x_1^3~~x_1^2x_2~~x_1x_2^2~~\nonumber \\
x_2^3~~x_1^4~~x_1^3x_2~~x_1^2x_2^2~~x_1x_2^3~~x_1^4~~ \nonumber \\
x_1\mu ~~x_2\mu ~~x_1^2\mu ~~  x_1x_2\mu ~~ x_2^2\mu ~~ \mu^2]^T
\end{eqnarray}
of size $ L=18 $, and the weighted basic function vector as $ \mathcal{W}_L =\Psi_L $.
According to the policy improvement rule \eqref{eq_3.11}, the iterative control policy is $ u^{(i)}(x) = K^{(i)}[x_1 ~~x_2 ~~x_1^2 ~~  x_1x_2 ~~ x_2^2]^T $ with control gain $ K^{(i)} = [k_1^{(i)}~~k_2^{(i)}~~k_3^{(i)}~~k_4^{(i)}~~k_5^{(i)}] $ $ = - 0.5 (\theta_{18}^{(i)})^{-1} [\theta_{13}^{(i)}~~\theta_{14}^{(i)}~~\theta_{15}^{(i)}~~\theta_{16}^{(i)}~~\theta_{17}^{(i)}] $. Consider that the optimal control policy is  $ u^*(x) = -x_1x_2 $, which can be represented as $ u^*(x) = K[x_1 ~~x_2 ~~x_1^2 ~~  x_1x_2 ~~ x_2^2]^T $ with the idea optimal control gain $ K = [0 ~~0 ~~0~~ -1~~ 0] $.

\begin{figure}
\centering	\includegraphics[width=3.2in]{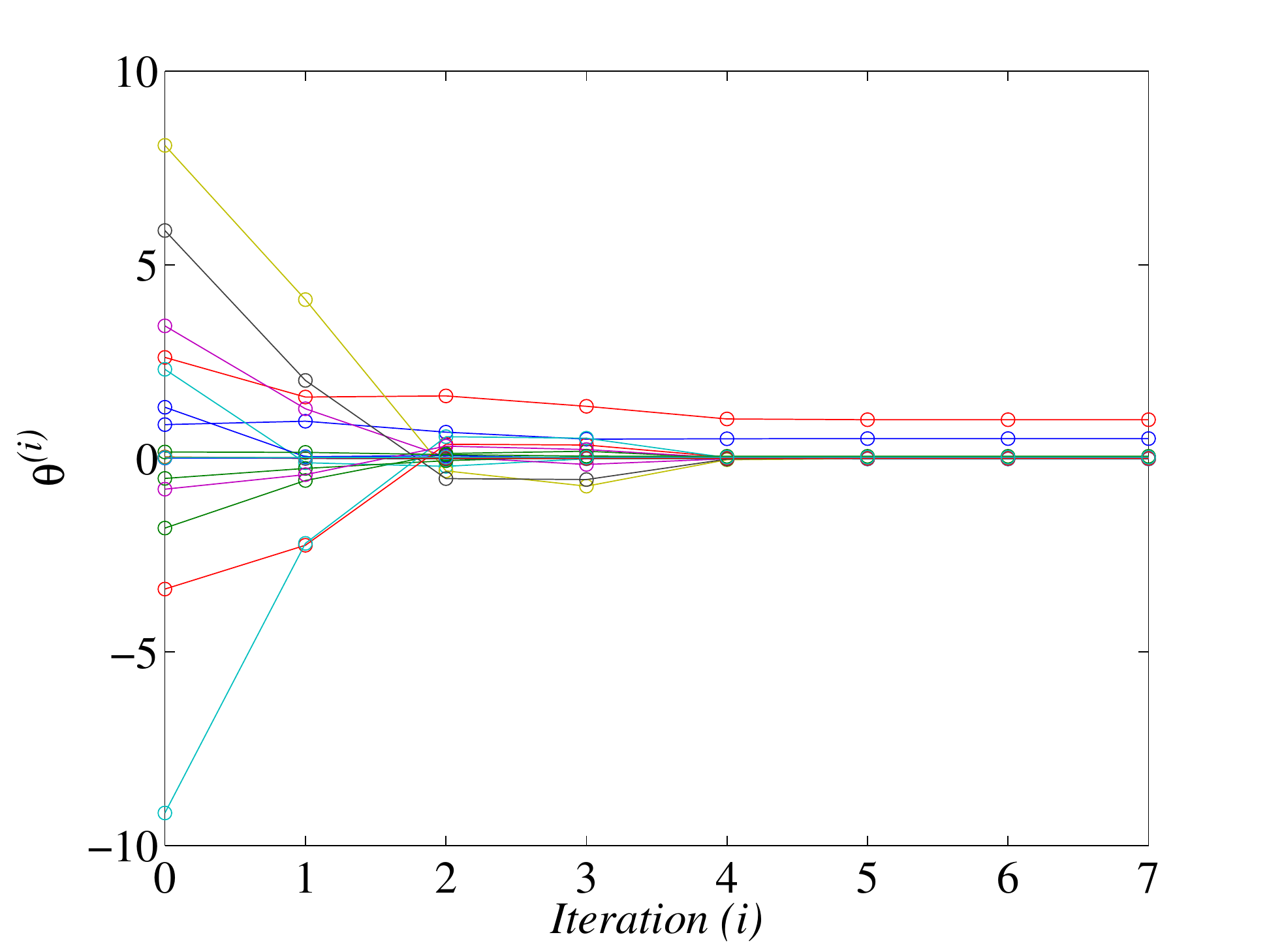}
	\caption{For example 2, all parameters of vector $ \theta^{(i)} $ obtained by PIQL algorithm.}
		\label{fig7}
\end{figure}

\begin{figure}
\centering	\includegraphics[width=3.2in]{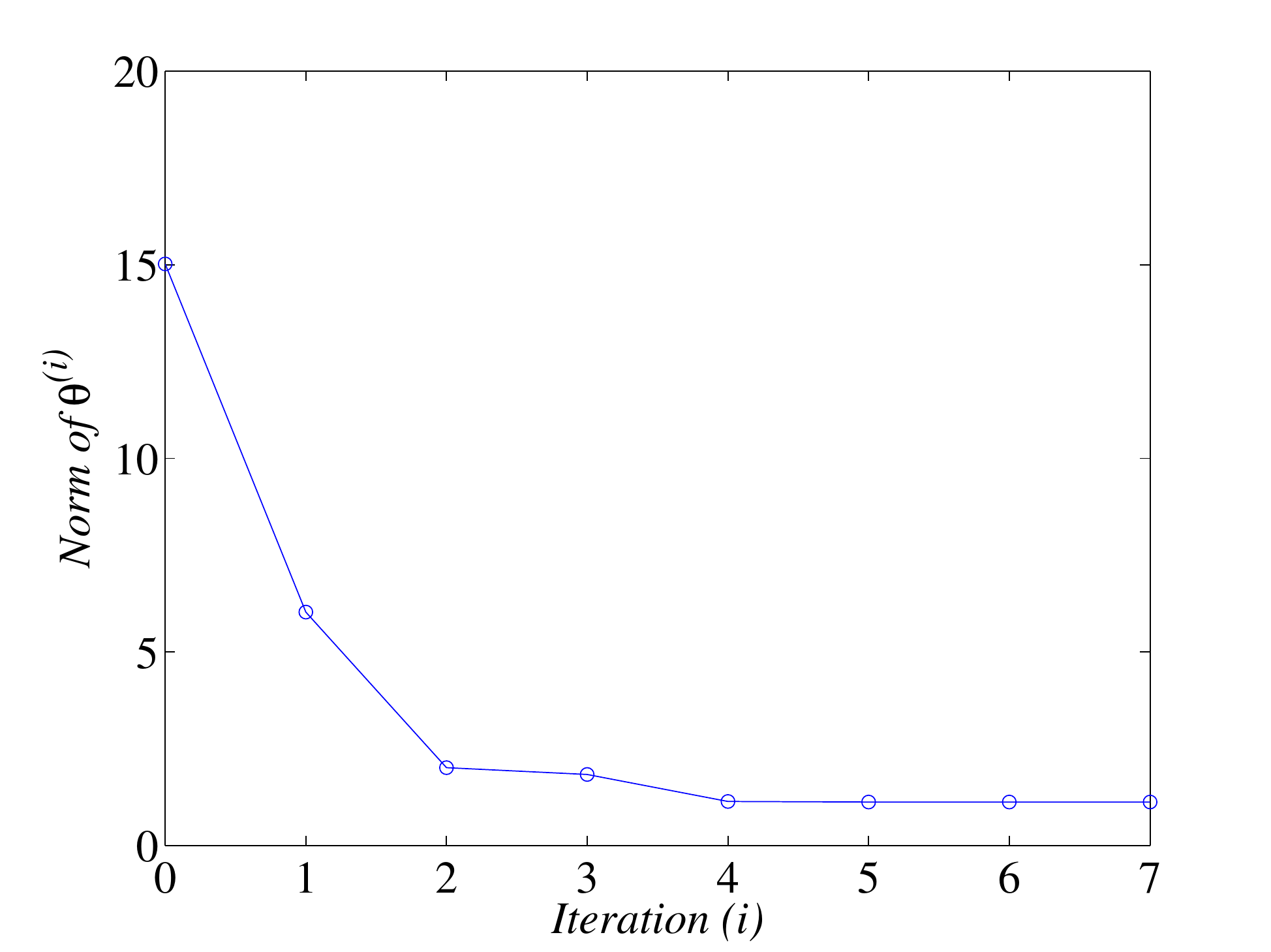}
	\caption{For example 2, the norm $ \Vert \theta^{(i)} \Vert $ obtained by PIQL algorithm.}
		\label{fig8}
\end{figure}

\begin{figure}
\centering	\includegraphics[width=3.2in]{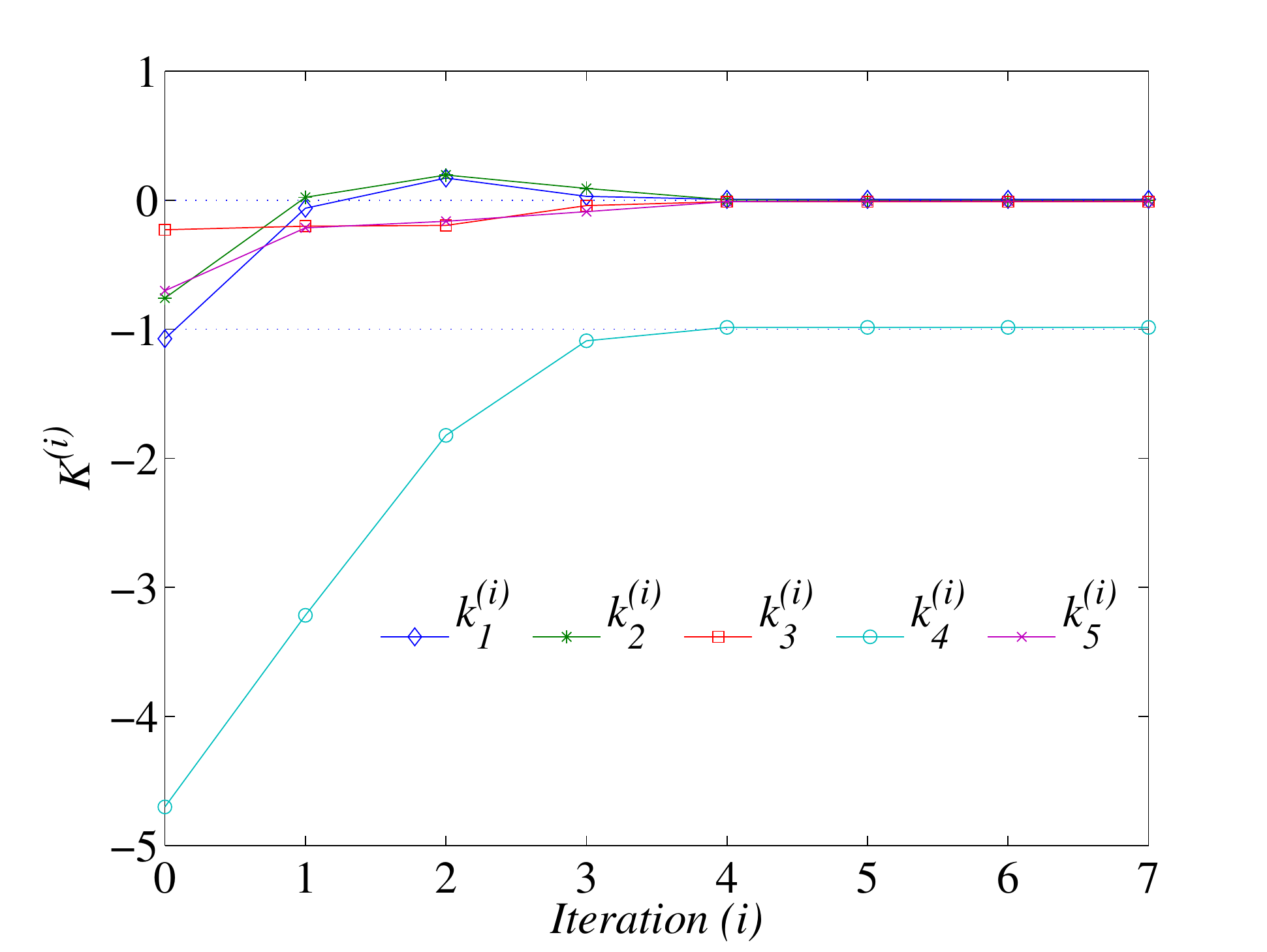}
	\caption{For example 2, the iterative control gain $ K^{(i)} $ obtained by PIQL algorithm.}
		\label{fig9}
\end{figure}

\begin{figure}
\centering	\includegraphics[width=3.2in]{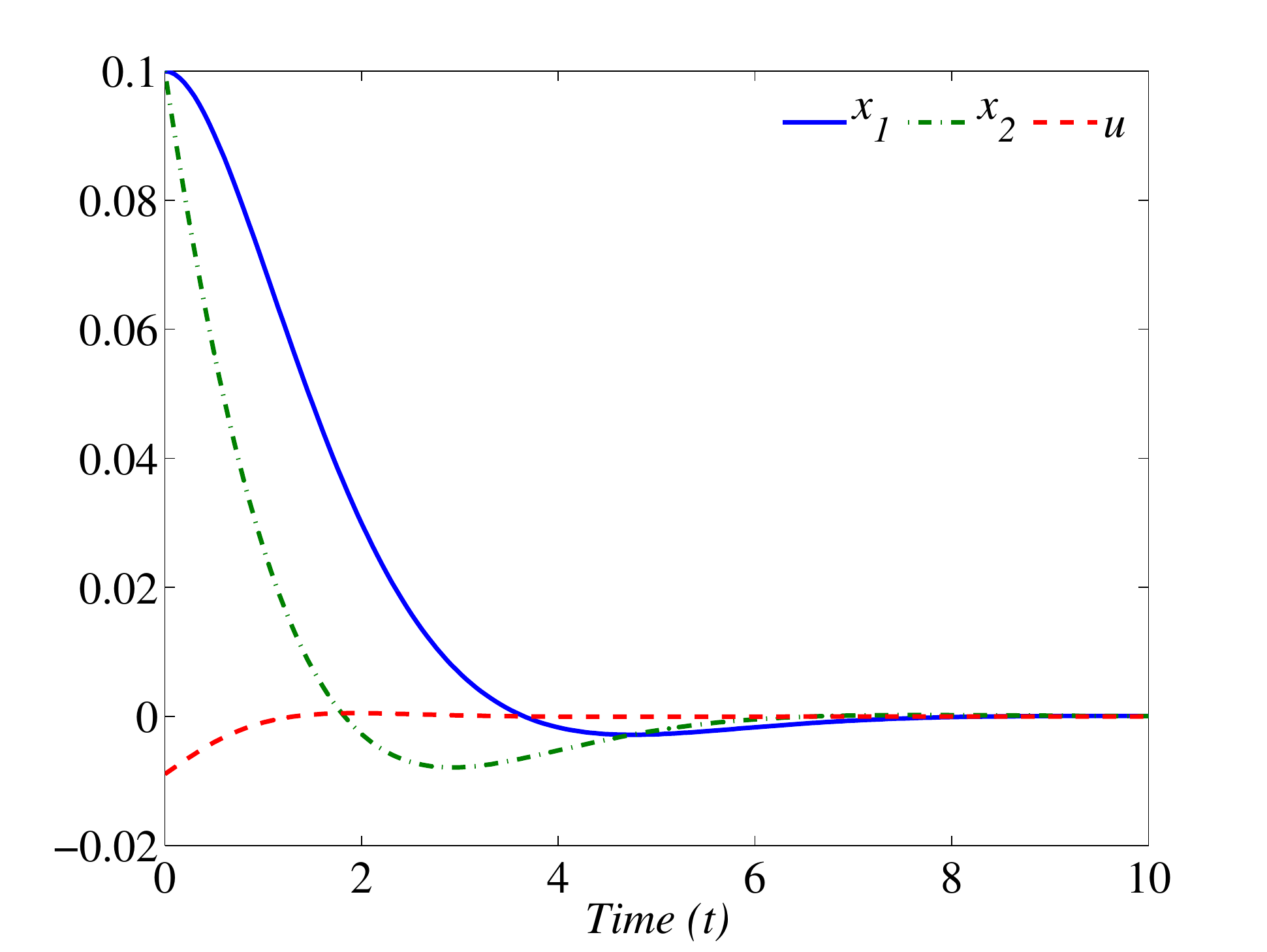}
	\caption{For example 2, the closed-loop trajectories of system state $ x(t) $ and control signal $ u(t) $ obtained by PIQL algorithm.}
		\label{fig10}
\end{figure}

First, the PIQL algorithm (i.e., Algorithm \ref{algorithm_3.2}) is used to solve the model-free optimal control problem of the system \eqref{eq_6.2}. It is observed that the algorithm achieves convergence at $ i = 7$ iteration, where the parameter vector $ \theta^{(i)} $ converges to
\begin{eqnarray}
\theta^{(7)} = [
    0.5097~~
    0.0078~~
    0.9977~~
   -0.0189~~
   -0.0142~~\nonumber \\
   -0.0060~~
    0.0120~~
    0.0091~~
    0.0038~~
    0.0215~~\nonumber \\
   -0.0057~~
   -0.0078~~
   -0.0003~~
    0.0000~~
    0.0006~~\nonumber \\
    0.0497~~
    0.0004~~
    0.0252]^T.\nonumber
\end{eqnarray}
and the iterative control gain $ K^{(i)} $ converges to
\begin{eqnarray}
K^{(7)} = [0.0067~~-0.0009~~-0.0112~~-0.9860~~-0.0076]. \nonumber
\end{eqnarray}
Figure \ref{fig7}-\ref{fig8} give the the parameter vector $ \theta^{(i)} $ and the norm $ \Vert \theta^{(i)} \Vert $ at each iteration, respectively. Figure \ref{fig9} shows the control gain $ K^{(i)} $ at each iteration, where the dot lines represent idea value of the optimal control gain $ K $. It is noted that good convergence is achieved. Then, the convergent control policy $ \hat{u}^{(7)}(x) $ is used for real control of the system \eqref{eq_6.2}. It is observed that the cost is $ 0.0150 $. Figure \ref{fig10} shows the closed-loop trajectories of system state and control signal.

\begin{figure}
\centering	\includegraphics[width=3.2in]{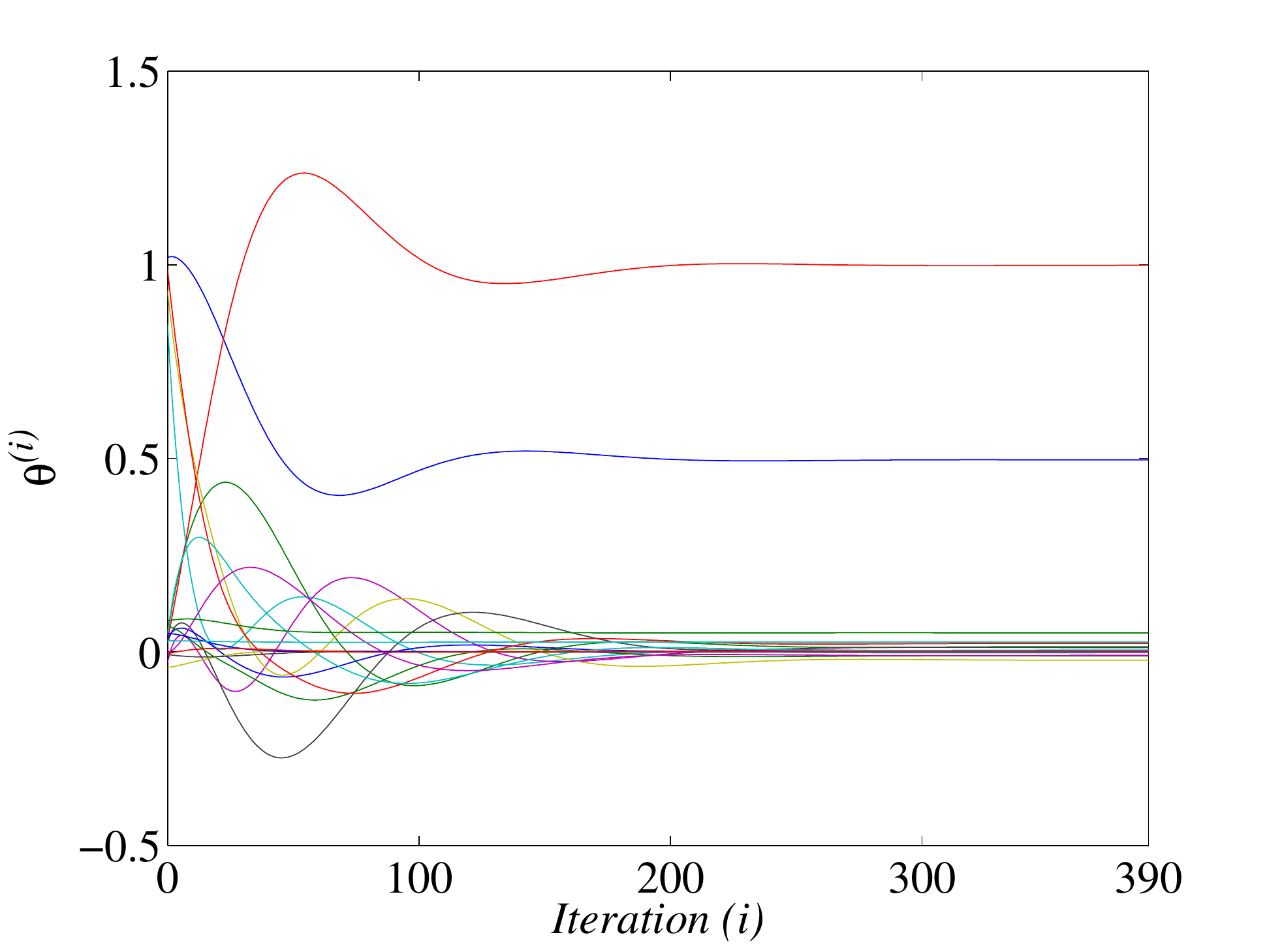}
	\caption{For example 2, all parameters of vector $ \theta^{(i)} $ obtained by VIQL algorithm.}
		\label{fig11}
\end{figure}

\begin{figure}
\centering	\includegraphics[width=3.2in]{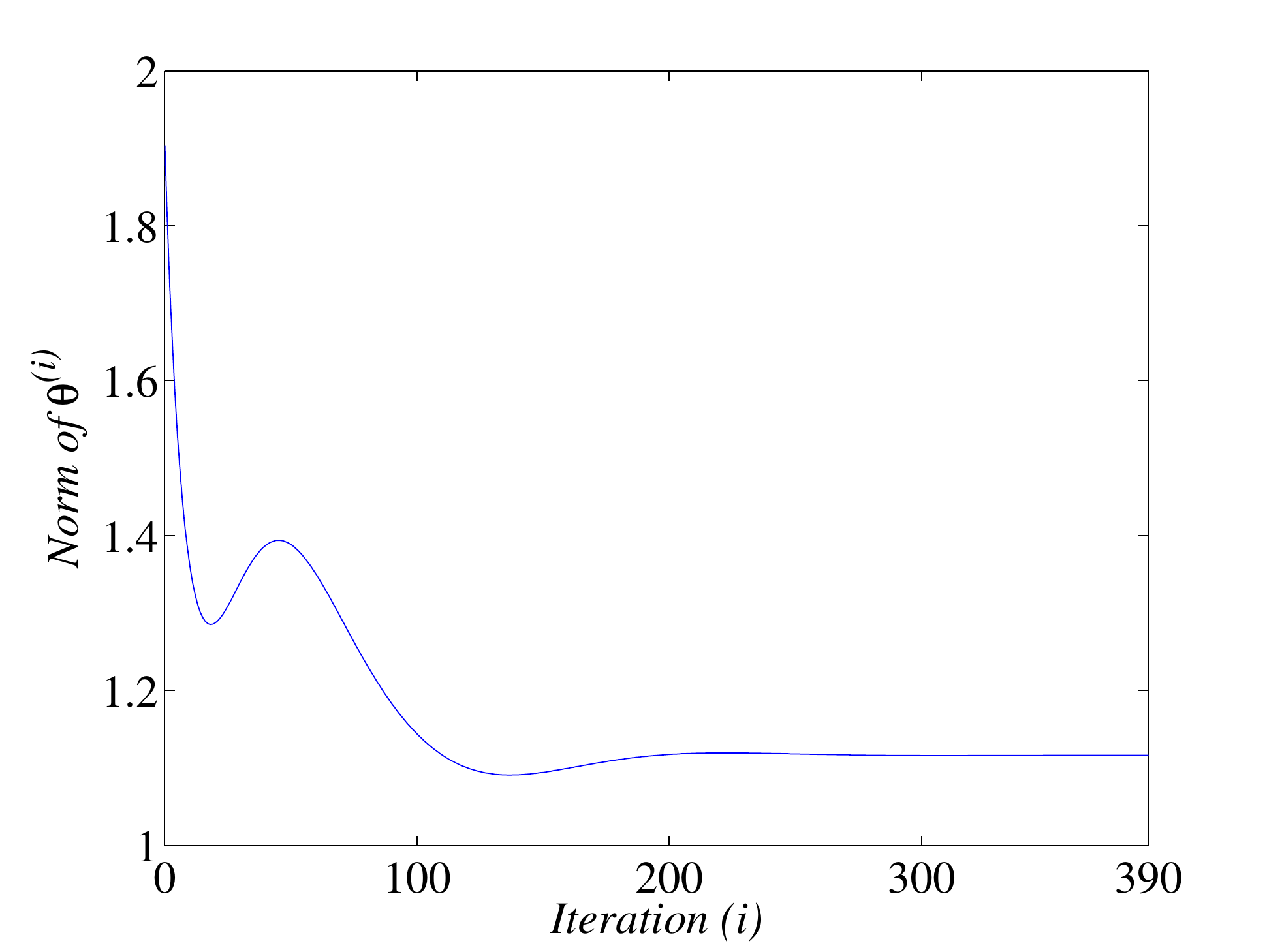}
	\caption{For example 2, the norm $ \Vert \theta^{(i)} \Vert $ obtained by VIQL algorithm.}
		\label{fig12}
\end{figure}

\begin{figure}
\centering	\includegraphics[width=3.2in]{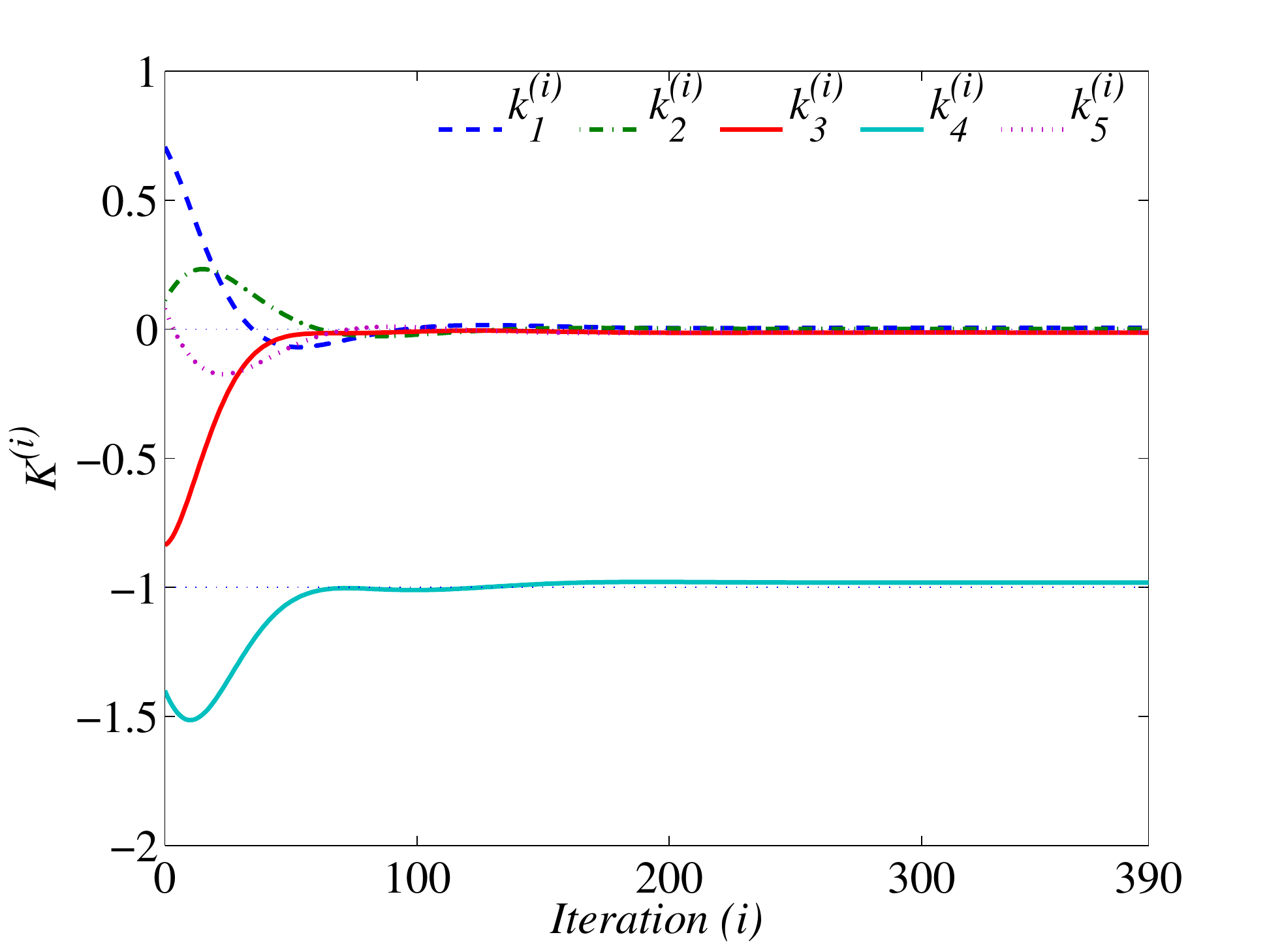}
	\caption{For example 2, the iterative control gain $ K^{(i)} $ obtained by VIQL algorithm.}
		\label{fig13}
\end{figure}

Next, by using the same basic function vector \eqref{eq_6.3}, the VIQL algorithm (i.e., Algorithm \ref{algorithm_4.2}) is employed to solve the model-free optimal control problem of system \eqref{eq_6.2}. Figures \ref{fig11}-\ref{fig13} give the simulation results, where the VIQL algorithm achieves convergence at $ i = 390 $ iteration. Figures \ref{fig11} and \ref{fig12} show the parameter vector $ \theta^{(i)} $ and its norm $ \Vert \theta^{(i)} \Vert $ respectively, where the parameter vector converges to
\begin{eqnarray}
\theta^{(390)} = [
    0.4958~~
    0.0132~~
    0.9983~~
   -0.0021~~
    0.0022~~\nonumber \\
   -0.0214~~
    0.0112~~
    0.0029~~
   -0.0102~~
    0.0219~~\nonumber \\
    0.0047~~
   -0.0089~~
   -0.0003~~
   -0.0001~~
    0.0007~~\nonumber \\
    0.0495~~
    0.0006~~
    0.0252]^T.\nonumber
\end{eqnarray}
Figure \ref{fig13} shows the control gain $ K^{(i)} $ at each iteration, where the dot lines represent idea value of the optimal control gain $ K $. It is observed that $ K^{(i)} $ converges to
\begin{eqnarray}
K^{(390)} = [0.0055~~0.0016~~-0.0130~~-0.9813~~-0.0117]\nonumber
\end{eqnarray}
which approaches to the optimal control gain $ K $. With the convergent control policy $ \hat{u}^{(390)}(x) $, closed-loop simulation is conducted on the real system \eqref{eq_6.2}. It is observed that the cost is $ 0.0150 $, and the the closed-loop trajectories of system state and control signal are almost the same as that in Figure \ref{fig10}, which is omitted here.

\section{Conclusions} \label{Sec_7}
In this paper, the model-free optimal control problem of general nonlinear continuous-time systems is considered, and two QL methods are developed.
First, PIQL and VIQL algorithms are proposed and their convergence is proved. For implementation purpose, the MWR is employed to derive a update law for unknown parameters.  Both PIQL and VIQL algorithms are model-free off-policy RL methods, which learn the optimal control policy offline from real system data and then be used for real-time control. Subsequently, PIQL and VIQL algorithms are simplified to solve the LQR problem of linear systems. Finally, a linear F-16 aircraft plant and a numerical nonlinear system are used to test the developed QL algorithms, and the obtained simulation results demonstrate their effectiveness.

\ifCLASSOPTIONcaptionsoff
  \newpage
\fi



%
\renewcommand\refname{References}
\bibliographystyle{IEEEtr}
\bibliography{References}

\begin{thebibliography}{10}

\bibitem{bertsekas1996neuro}
D.~P. Bertsekas and J.~N. Tsitsiklis, {\em Neuro-Dynamic Programming}.
\newblock Belmont, Mass.: Athena Scientific, 1996.

\bibitem{sutton1998reinforcement}
R.~S. Sutton and A.~G. Barto, {\em Reinforcement Learning: An Introduction}.
\newblock Cambridge Univ Press, Massachusetts London, England, 1998.

\bibitem{kaelbling1996reinforcement}
L.~P. Kaelbling, M.~L. Littman, and A.~W. Moore, ``Reinforcement learning: A
  survey,'' {\em Journal of Artificial Intelligence Research}, vol.~4,
  pp.~237--285, 1996.

\bibitem{gosavi2009reinforcement}
A.~Gosavi, ``Reinforcement learning: A tutorial survey and recent advances,''
  {\em INFORMS Journal on Computing}, vol.~21, no.~2, pp.~178--192, 2009.

\bibitem{watkins1989learning}
C.~J. C.~H. Watkins, {\em Learning from delayed rewards}.
\newblock PhD thesis, University of Cambridge, 1989.

\bibitem{watkins1992q}
C.~J. C.~H. Watkins and P.~Dayan, ``Q-learning,'' {\em Machine learning},
  vol.~8, no.~3-4, pp.~279--292, 1992.

\bibitem{tsitsiklis1994asynchronous}
J.~N. Tsitsiklis, ``Asynchronous stochastic approximation and {Q}-learning,''
  {\em Machine Learning}, vol.~16, no.~3, pp.~185--202, 1994.

\bibitem{sutton2014new}
R.~S. Sutton, A.~R. Mahmood, D.~Precup, M.~CA, H.~van Hasselt, and U.~CA, ``A
  new {Q}($\lambda$) with interim forward view and {Monte Carlo} equivalence,''
  in {\em Proceedings of the 31st International Conference on Machine
  Learning}, 2014.

\bibitem{tsitsiklis1997analysis}
J.~N. Tsitsiklis and B.~Van~Roy, ``An analysis of temporal-difference learning
  with function approximation,'' {\em IEEE Transactions on Automatic Control},
  vol.~42, no.~5, pp.~674--690, 1997.

\bibitem{jaakkola1994convergence}
T.~Jaakkola, M.~I. Jordan, and S.~P. Singh, ``On the convergence of stochastic
  iterative dynamic programming algorithms,'' {\em Neural computation}, vol.~6,
  no.~6, pp.~1185--1201, 1994.

\bibitem{maei2010gq}
H.~R. Maei and R.~S. Sutton, ``{GQ} ($\lambda$): A general gradient algorithm
  for temporal-difference prediction learning with eligibility traces,'' in
  {\em Proceedings of the Third Conference on Artificial General Intelligence},
  vol.~1, pp.~91--96, 2010.

\bibitem{Even-Dar2003learning}
E.~Even-Dar and Y.~Mansour, ``Learning rates for {Q}-learning,'' {\em Journal
  of Machine Learning Research}, vol.~5, pp.~1--25, Dec. 2003.

\bibitem{gosavi2004reinforcement}
A.~Gosavi, ``A reinforcement learning algorithm based on policy iteration for
  average reward: Empirical results with yield management and convergence
  analysis,'' {\em Machine Learning}, vol.~55, no.~1, pp.~5--29, 2004.

\bibitem{peng1996incremental}
J.~Peng and R.~J. Williams, ``Incremental multi-step {Q}-learning,'' {\em
  Machine Learning}, vol.~22, no.~1-3, pp.~283--290, 1996.

\bibitem{bhatnagar2008new}
S.~Bhatnagar and K.~M. Babu, ``New algorithms of the {Q}-learning type,'' {\em
  Automatica}, vol.~44, no.~4, pp.~1111--1119, 2008.

\bibitem{hull2003optimal}
D.~G. Hull, {\em Optimal Control Theory for Applications}.
\newblock Troy, NY: Springer, 2003.

\bibitem{bertsekas2005dynamic}
D.~P. Bertsekas, {\em Dynamic Programming and Optimal Control}, vol.~1.
\newblock Nashua: Athena Scientific, 2005.

\bibitem{lewis2013optimal}
F.~L. Lewis, D.~Vrabie, and V.~L. Syrmos, {\em Optimal Control}.
\newblock Hoboken, New Jersey: John Wiley \& Sons, Inc., 2013.

\bibitem{lewis2013reinforcement}
F.~L. Lewis and D.~Liu, {\em Reinforcement Learning and Approximate Dynamic
  Programming for Feedback Control}, vol.~17.
\newblock Hoboken, New Jersey: John Wiley \& Sons, Inc., 2013.

\bibitem{lewis2012reinforcement}
F.~L. Lewis, D.~Vrabie, and K.~G. Vamvoudakis, ``Reinforcement learning and
  feedback control: Using natural decision methods to design optimal adaptive
  controllers,'' {\em IEEE Control Systems}, vol.~32, no.~6, pp.~76--105, 2012.

\bibitem{vrabie2009neural}
D.~Vrabie and F.~L. Lewis, ``Neural network approach to continuous-time direct
  adaptive optimal control for partially unknown nonlinear systems,'' {\em
  Neural Networks}, vol.~22, no.~3, pp.~237--246, 2009.

\bibitem{zhang2011data}
H.~Zhang, L.~Cui, X.~Zhang, and Y.~Luo, ``Data-driven robust approximate
  optimal tracking control for unknown general nonlinear systems using adaptive
  dynamic programming method,'' {\em IEEE Transactions on Neural Networks},
  vol.~22, no.~12, pp.~2226--2236, 2011.

\bibitem{liu2012neural}
D.~Liu, D.~Wang, D.~Zhao, Q.~Wei, and N.~Jin, ``Neural-network-based optimal
  control for a class of unknown discrete-time nonlinear systems using
  globalized dual heuristic programming,'' {\em IEEE Transactions on Automation
  Science and Engineering}, vol.~9, no.~3, pp.~628--634, 2012.

\bibitem{jiang2012computational}
Y.~Jiang and Z.-P. Jiang, ``Computational adaptive optimal control for
  continuous-time linear systems with completely unknown dynamics,'' {\em
  Automatica}, vol.~48, no.~10, pp.~2699--2704, 2012.

\bibitem{dierks2012online}
T.~Dierks and S.~Jagannathan, ``Online optimal control of affine nonlinear
  discrete-time systems with unknown internal dynamics by using time-based
  policy update,'' {\em IEEE Transactions on Neural Networks and Learning
  Systems}, vol.~23, no.~7, pp.~1118--1129, 2012.

\bibitem{wang2012optimal}
D.~Wang, D.~Liu, Q.~Wei, D.~Zhao, and N.~Jin, ``Optimal control of unknown
  nonaffine nonlinear discrete-time systems based on adaptive dynamic
  programming,'' {\em Automatica}, vol.~48, no.~8, pp.~1825--1832, 2012.

\bibitem{modaresadaptive2013adaptive}
H.~Modares, F.~L. Lewis, and M.-B. Naghibi-Sistani, ``Adaptive optimal control
  of unknown constrained-input systems using policy iteration and neural
  networks,'' {\em IEEE Transactions on Neural Networks and Learning Systems},
  vol.~24, no.~10, pp.~1513--1525, 2013.

\bibitem{modares2014integral}
H.~Modares, F.~L. Lewis, and M.-B. Naghibi-Sistani, ``Integral reinforcement
  learning and experience replay for adaptive optimal control of
  partially-unknown constrained-input continuous-time systems,'' {\em
  Automatica}, vol.~50, no.~1, pp.~193--202, 2014.

\bibitem{wei2014adaptive}
Q.~Wei and D.~Liu, ``Adaptive dynamic programming for optimal tracking control
  of unknown nonlinear systems with application to coal gasification,'' {\em
  IEEE Transactions on Automation Science and Engineering}, p.~In Press, 2014.

\bibitem{li2014integral}
H.~Li, D.~Wang, and D.~Liu, ``Integral reinforcement learning for linear
  continuous-time zero-zum games with completely unknown dynamics,'' {\em IEEE
  Transactions on Automation Science and Engineering}, vol.~11, no.~3,
  pp.~706--714, 2014.

\bibitem{modares2014linear}
H.~Modares and F.~L. Lewis, ``Linear quadratic tracking control of
  partially-unknown continuous-time systems using reinforcement learning,''
  {\em IEEE Transactions on Automatic Control}, p.~In Press, 2014.

\bibitem{yang2014reinforcement}
X.~Yang, D.~Liu, and D.~Wang, ``Reinforcement learning for adaptive optimal
  control of unknown continuous-time nonlinear systems with input
  constraints,'' {\em International Journal of Control}, vol.~87, no.~3,
  pp.~553--566, 2014.

\bibitem{al2007model}
A.~Al-Tamimi, F.~L. Lewis, and M.~Abu-Khalaf, ``Model-free {Q}-learning designs
  for linear discrete-time zero-sum games with application to {H}-infinity
  control,'' {\em Automatica}, vol.~43, no.~3, pp.~473--481, 2007.

\bibitem{kim2010model}
J.-H. Kim and F.~L. Lewis, ``Model-free {$H_\infty$} control design for unknown
  linear discrete-time systems via {Q}-learning with {LMI},'' {\em Automatica},
  vol.~46, no.~8, pp.~1320--1326, 2010.

\bibitem{lee2012integral}
J.~Y. Lee, J.~B. Park, and Y.~H. Choi, ``Integral {Q}-learning and explorized
  policy iteration for adaptive optimal control of continuous-time linear
  systems,'' {\em Automatica}, vol.~48, no.~11, pp.~2850--2859, 2012.

\bibitem{palanisamy2014continuous}
M.~Palanisamy, H.~Modares, F.~Lewis, and M.~Aurangzeb, ``Continuous-time
  {Q}-learning for infinite-horizon discounted cost linear quadratic regulator
  problems,'' {\em IEEE Transactions on Cybernetics}, p.~In Press, 2014.

\bibitem{kiumarsi2014reinforcement}
B.~Kiumarsi, F.~L. Lewis, H.~Modares, A.~Karimpour, and M.-B. Naghibi-Sistani,
  ``Reinforcement {Q}-learning for optimal tracking control of linear
  discrete-time systems with unknown dynamics,'' {\em Automatica}, vol.~50,
  no.~4, pp.~1167--1175, 2014.

\bibitem{precup2001off}
D.~Precup, R.~S. Sutton, and S.~Dasgupta, ``Off-policy temporal-difference
  learning with function approximation,'' in {\em Proceedings of the 18th
  International Conference on Machine Learning}, pp.~417--424, 2001.

\bibitem{luo2014off}
B.~Luo, H.-N. Wu, and T.~Huang, ``Off-policy reinforcement learning for {$
  H_\infty $} control design,'' {\em IEEE Transactions on Cybernetics},
  vol.~DOI: 10.1109/TCYB. 2014.2319577, p.~In Press, 2014.

\bibitem{peter1978new}
G.~Peter~Lepage, ``A new algorithm for adaptive multidimensional integration,''
  {\em Journal of Computational Physics}, vol.~27, no.~2, pp.~192--203, 1978.

\bibitem{stevens2003aircraft}
B.~L. Stevens and F.~L. Lewis, {\em Aircraft Control and Simulation}.
\newblock Wiley-Interscience, 2003.

\bibitem{vamvoudakis2014online}
K.~G. Vamvoudakis, D.~Vrabie, and F.~L. Lewis, ``Online adaptive algorithm for
  optimal control with integral reinforcement learning,'' {\em International
  Journal of Robust and Nonlinear Control}, p.~In Press, 2014.

\bibitem{vamvoudakis2010online}
K.~G. Vamvoudakis and F.~L. Lewis, ``Online actor--critic algorithm to solve
  the continuous-time infinite horizon optimal control problem,'' {\em
  Automatica}, vol.~46, no.~5, pp.~878--888, 2010.

\bibitem{nevistic1996optimality}
V.~Nevisti{\'c} and J.~A. Primbs, ``Optimality of nonlinear design techniques:
  a converse {HJB} approach,'' tech. rep., California Institute of Technology,
  TR96-022, 1996.

\end{thebibliography}

%
%

%
%
%
%
%




\end{document}